\documentclass[a4paper,UKenglish,cleveref, autoref, thm-restate]{lipics-v2021}

\hideLIPIcs  

\title{An Improved Approximation Algorithm for the Max-\texorpdfstring{$3$}{3}-Section Problem} 

\titlerunning{An Improved Approximation Algorithm for the Max-$3$-Section Problem} 

\author{Dor Katzelnick}{The Henry and Marilyn Taub Faculty of Computer Science, Technion, Israel}{dkatzelnick@cs.technion.ac.il}{https://orcid.org/0009-0006-7482-5146}{}

\author{Aditya Pillai}{H. Milton Stewart School of Industrial and Systems Engineering, Georgia Institute of Technology, Atlanta, USA}{apillai32@gatech.edu}{https://orcid.org/0009-0006-2691-8313}{}

\author{Roy Schwartz}{The Henry and Marilyn Taub Faculty of Computer Science, Technion, Israel}{schwartz@cs.technion.ac.il}{}{}

\author{Mohit Singh}{H. Milton Stewart School of Industrial and Systems Engineering, Georgia Institute of Technology, Atlanta, USA}{mohit.singh@isye.gatech.edu}{https://orcid.org/0000-0002-0827-233X}{}

\authorrunning{D. Katzelnick, A. Pillai, R. Schwartz, and M. Singh}

\Copyright{Dor Katzelnick, Aditya Pillai, Roy Schwartz, and Mohit Singh}



\category{} 

\funding{Mohit Singh and Aditya Pillai were supported in part by NSF CCF-2106444 and NSF CCF-1910423.
Dor Katzelnick and Roy Schwartz receive funding from the European Union’s Horizon 2020 research and innovation program under
grant agreement no. 852870-ERC-SUBMODULAR.}

\acknowledgements{The authors would like to thank Uri Zwick for insightful discussions.}

\nolinenumbers 

\EventEditors{Inge Li G{\o}rtz, Martin Farach-Colton, Simon J. Puglisi, and Grzegorz Herman}
\EventNoEds{4}
\EventLongTitle{31st Annual European Symposium on Algorithms (ESA 2023)}
\EventShortTitle{ESA 2023}
\EventAcronym{ESA}
\EventYear{2023}
\EventDate{September 4--6, 2023}
\EventLocation{Amsterdam, the Netherlands}
\EventLogo{}
\SeriesVolume{274}
\ArticleNo{82}

\usepackage{xspace}
\usepackage{xcolor}
\usepackage{amsfonts}
\usepackage{amsmath}
\usepackage{amssymb}
\usepackage{amsthm}
\usepackage{nicefrac}
\usepackage[linesnumbered,ruled,vlined]{algorithm2e}

\newcommand{\RR}{\mathbb{R}}

\def \C {\mathcal{C}}
\def \RR {\mathbb{R}}

\def \EE {\mathbb{E}}

\newcommand{\maxcut}{\textsc{{Max}-\textsc{Cut}}\xspace}
\newcommand{\maxbis}{\textsc{Max}-\textsc{Bisection}\xspace}
\newcommand{\maxkcut}{\textsc{Max}-$k$-\textsc{Cut}\xspace}
\newcommand{\maxtricut}{\textsc{Max}-$3$-\textsc{Cut}\xspace}
\newcommand{\maxksec}{\textsc{Max}-$k$-\textsc{Section}\xspace}
\newcommand{\maxtris}{\textsc{Max}-$3$-\textsc{Section}\xspace}

\newcommand{\by}{\mathbf{y}}
\newcommand{\bg}{\mathbf{g}}
\newcommand{\bz}{\mathbf{z}}
\newcommand{\bc}{\mathbf{c}}

\begin{document}

\maketitle

\begin{abstract}
We consider the \maxtris problem, where we are given an undirected graph $ G=(V,E)$ equipped with non-negative edge weights $w :E\rightarrow \mathbb{R}_+$ and the goal is to find a partition of $V$ into three equisized parts while maximizing the total weight of edges crossing between different parts.
\maxtris is closely related to other well-studied graph partitioning problems, {\em e.g.}, \maxcut, \maxtricut, and \maxbis.
We present a polynomial time algorithm achieving an approximation of $ 0.795$, that improves upon the previous best known approximation of $ 0.673$.
The requirement of multiple parts that have equal sizes renders \maxtris much harder to cope with compared to, {\em e.g.}, \maxbis.
We show a new algorithm that combines the existing approach of Lassere hierarchy along with a random cut strategy that suffices to give our result.
\end{abstract}

\section{Introduction}\label{sec:intro}
In this paper we study the \maxtris problem: given an undirected graph $G=(V,E)$ equipped with non-negative edge weights $w:E\rightarrow \RR_+$ the goal is to partition the vertex set $V$ into three equisized parts while maximizing the total weight of edges that cross between different parts. 
\maxtris is closely related to other classic graph partitioning problems, where given the same input as in \maxtris the goal is to output a partition of the vertex set $V$ (possibly given some problem specific constraint) while maximizing the total weight of edges that cross between different parts.
Perhaps the most famous of these problems is \maxcut, whose constraint is that the partition contains only two parts with no restriction on the size of these parts.
\maxcut is one of Karp's 21 NP-hard problems \cite{karp1972reducibility} and in their seminal work Goemans and Williamson \cite{GW95} presented an approximation of $ 0.8786$ using semi-definite programming and random hyperplane rounding.
It is known that the latter result is tight, assuming the unique games conjecture, as proved by Khot, Kindler, Mossel, and O'Donnell \cite{KKMO}.


A natural problem that generalizes \maxcut is \maxkcut, whose constraint is that the partition contains $k$ parts with no restriction on the size of these parts.
A simple algorithm that returns a uniform random solution, {\em i.e.}, every vertex is assigned independently and uniformly to one of the $k$ parts, achieves an approximation of $ 1-\nicefrac{1}{k}$.
Several works aimed at improving the guarantee of this simple algorithm, {\em e.g.}, \cite{frieze1997improved, goemans2001approximation, de2004approximate, Newman2018ComplexSP}.
For example, a notable case that attracted much attention is the \maxtricut problem \cite{frieze1997improved, goemans2001approximation, de2004approximate, Newman2018ComplexSP}, whose best approximation is $ 0.836$ and was given by Goemans and Williamson \cite{goemans2001approximation}.
Interestingly, the latter guarantee is worse than the approximation known for \maxcut.
For general values of $k$, it was shown by Frieze and Jerrum \cite{frieze1997improved} that an approximation of $ 1-\nicefrac{1}{k}+\Theta(\ln{k}/k^2)$ can be achieved.
Further improvements for the approximation guarantee were presented by de Klerk, Pasechnik and Warners \cite{de2004approximate}.


Adding a single global constraint to \maxcut that requires both parts to be of equal size leads to the classic problem of \maxbis.
\maxbis was extensively studied throughout the years, {\em e.g.}, \cite{frieze1997improved,Ye2001,halperin2002unified,feige2006rpr2,RaghavendraT12,AustrinBG16}.
This sequence of works currently culminates with the works of Raghavendra and Tan \cite{RaghavendraT12}, who present an approximation of $0.85$ that is based on rounding a Lasserre hierarchy semi-definite program, which was subsequently improved to $ 0.877$ by Austrin, Benabbas, and Georgiou \cite{AustrinBG16} who improved the former rounding.
The question whether one can obtain for \maxbis the same approximation guarantee that is known for \maxcut  remains a tantalizing open problem.


Both \maxtris and \maxbis are captured by the \maxksec problem \cite{andersson2002approximation, gaur2008capacitated, ling2009approximation,de2012semidefinite}, which falls in the above broad family of graph partitioning problems and whose constraint is that the partition contains $k$ equisized parts.
Similarly to \maxkcut, it is known that the simple algorithm that returns a uniform random solution achieves an approximation of $ 1-\nicefrac{1}{k}$.
Thus, it is no surprise that the goal of past research, {\em e.g.}, \cite{andersson2002approximation, gaur2008capacitated, ling2009approximation}, was to improve upon this guarantee.
For the special case of \maxtris, Ling \cite{ling2009approximation} presented an approximation of $0.6733$ that is based on rounding a semi-definite programming relaxation similarly to \maxtricut \cite{goemans2001approximation}.
For general values of $k$, Andersson \cite{andersson2002approximation}  presented an approximation of $ 1-\nicefrac{1}{k}+\Theta(1/k^3)$, again by rounding a suitable semi-definite program relaxation.
Additionally, Gaur, Krishnamurti, and Kohli \cite{gaur2008capacitated} presented a local search algorithm for a more general problem in which each part has a (possibly different) limit on its size.


The main focus of this work is the \maxtris problem.
For $k=2$, the best known approximation for \maxbis (which is essentially \maxcut with a single global constraint on the size of the first part in the partition) equals $ 0.877$ and is (almost) identical to the best possible approximation of $0.878$ for \maxcut.
However, when $ k=3$, the best known approximation for \maxtris (which is essentially \maxtricut with two global constraints on the size of the first two pieces in the partition) equals $ 0.6733$ and is far from the best known approximation of $0.836$ for \maxtricut.
Moreover, the former approximation guarantee of $0.6733$ for \maxtris only slightly improves upon the $\nicefrac{2}{3}$ approximation guarantee of the trivial algorithm that simply returns a uniform random solution.
Thus, we aim to understand and minimize the gap that exists for $k=3$. 

\subsection{Our Results and Techniques}
We present the following main algorithmic result for the \maxtris problem:
\begin{theorem}\label{theorem:final}
There is a polynomial time algorithm for \maxtris that achieves an approximation guarantee of at least $ 0.795$.
\end{theorem}
It is important to note that the approximation guarantee of the above theorem improves upon the previous best known algorithm for \maxtris \cite{ling2009approximation} that achieves an approximation of $ 0.6773$.
As proving the exact approximation guarantee of our algorithm is an involved task (for reasons that will be clear soon), we also present the following conjecture that is based on {numerical evidence}:
\begin{conjecture}\label{conj:final}
The algorithm presented to prove Theorem \ref{theorem:final} achieves an approximation of $ 0.8192$ for \maxtris.    
\end{conjecture}

We further show how the algorithm we present for proving Theorem \ref{theorem:final} can be extended to general values of $k$.
First, we prove that the algorithm, alongside its analysis, cannot provide an improved approximation as $k$ increases (when compared to its approximation for the case of $k=3$).
Second, using numeric evidence we conjecture that the approximation of the algorithm remains $ 0.8192$ for $k$ ranging from $3$ up to $6$.
Hence, the above gives rise to the following extended conjecture:
\begin{conjecture}\label{conj:final2}
 There is a polynomial time algorithm achieving an approximation of $ 0.8192$ for \maxksec for $ k=3,4,5$.   
\end{conjecture}
The above conjecture provides, for $ k=4,5$, an improved approximation when compared to the previous best known result. The current best is a small improvement over the trivial randomized approximation algorithm by Andersson \cite{andersson2002approximation} which achieves an approximation of $1-\nicefrac{1}{k} + \Theta(k^{-3})$.  

When considering our approach to \maxtris we focus on two of its closely related problems: \maxtricut and \maxbis and examine the approaches used to design and analyze algorithms for both.
Let us start with \maxtricut.
The approach used to obtain the current best known algorithms for \maxtricut, {\em e.g.}, \cite{goemans2001approximation,de2004approximate,Newman2018ComplexSP}, is based on the random hyperplane rounding method due to Goemans and Williamson \cite{GW95} (as well as extensions of this method) of a semi-definite programming relaxation.
This approach, when applied to \maxtris, suffers a significant drawback since it does not preserve marginal values (a marginal value is the likelihood the relaxation assigns to the event that vertex $u$ belongs to part $i$).
Specifically, the expected number of vertices assigned to every part by the rounding algorithm might be incomparable to $ n/3$ and therefore applying these ideas as was done by Ling~\cite{ling2009approximation} leads only to a minor improvement over the random solution algorithm. 
It is worth noting that this drawback is already present when considering \maxbis. 

Let us now consider \maxbis, and specifically we focus on the approach of Raghavendra and Tan \cite{RaghavendraT12}  (as well as Austrin, Benabbas, and Georgiou \cite{AustrinBG16} who build upon \cite{RaghavendraT12}).
First, a Lasserre hierarchy of a natural semi-definite programming relaxation for \maxbis is solved.
Second, a rounding procedure that preserves the marginal values is applied to obtain a subset $ C_1\subseteq V$ of vertices such that: $(1)$ there are sufficiently many edges crossing the cut $C_1$ defines; and $(2)$ $C_1$ contains (roughly) $ n/2$ vertices.
Third, the solution is re-balanced to obtain a perfect bisection, {\em i.e.}, ensuring that $ |C_1|=n/2$ without a significant loss in the number of edges crossing between $C_1$ and $C_2=V\setminus C_1$.

There are two main difficulties when considering this approach in the context of \maxtris.
The first difficulty stems from the fact that we have three parts in \maxtris, whereas in \maxbis there are only two parts.
Therefore, if we use the above rounding procedure to find $ C_1\subseteq V$ of size (roughly) $ n/3$ with sufficiently many edges crossing the cut $C_1$ defines, it is not clear how to recurse and further partition $ V\setminus C_1$.
We note that in \maxbis no recursion is needed since $ C_2$ is chosen to be $ V\setminus C_1$.
However, in \maxtris $ V\setminus C_1$ still needs to be partitioned into $C_2$ and $C_3$.
Our solution to this difficulty is to {\em condition} the marginal values of vertices remaining in $ V\setminus C_1$ on the fact that each remaining vertex was not chosen to $C_1$.
Such a conditioning, intuitively, ensures that we preserve marginal values overall while recursing on $V\setminus C_1$. 

The second difficulty in applying this apporach arises from the following observation: we can show that if one first creates part $C_1$, and then creates part $C_2$ (and part $ C_3$ is all remaining vertices), then if the analysis is performed edge-by-edge (as is the case in both \cite{RaghavendraT12,AustrinBG16}) no approximation better than $ 0.7192$ can be achieved.
Specifically, we present a configuration of the vectors that correspond to the  endpoints of an edge that satisfy: $(1)$ the vectors are feasible for the semi-definite programming relaxation; and $(2)$ the ratio between the probability of this edge being separated by the rounding algorithm and the contribution of its vectors to the objective function of the relaxation is at most $ 0.7192$ (refer to Section \ref{sec:algorithm} for a formal definition of a configuration and to Observation \ref{obs:worstConfNoPerm} in Appendix \ref{app:Experiment} for the specific configuration).
An approximation of $ 0.7192$, if possible given the above approach, improves the current best known approximation of $ 0.6733$ \cite{ling2009approximation}.
However, we aim for a much larger improvement.
Our solution to this difficulty is to {\em uniformly permute} the order in which the parts are generated.
Since the approach based on \cite{RaghavendraT12,AustrinBG16} preserves marginals, this permutation allows us to better cope with problematic configurations.
It is important to note that a permutation is meaningless for \maxbis, since whether a vertex belongs to $C_1$ immediately implies whether it belongs to $C_2$ and vice versa.
In \maxtris this is obviously not the case.

Thus, following the above discussion, our approach builds upon the approach for \maxbis with two added ingredients.
The first is altering the marginal values the semi-definite programming relaxation provides via appropriate conditioning when recursing.
The second is uniformly permuting the order in which the rounding algorithm generates the parts.
We note that these two added ingredients introduce two main additional obstacles.
The first obstacle relates to the last re-balancing step.
In both \cite{RaghavendraT12,AustrinBG16} the re-balancing succeeds since it is proved that with a high probability each part by itself is close to being the desired size.
The method this is proved is by bounding the variance of the size of each part alone.
However, in our approach for \maxtris the bound on the variance of the size of a given part depends on the other parts as well.
This introduces technical issues and hence bounding of the variance requires much care.
The second obstacle relates to the computer assisted proof via {\em{branch and bound}} method  we employ in order to lower bound the performance ratio of our algorithm.
The expression of the separation probability of an edge by the rounding algorithm is involved, as both marginal values are altered when recursing and we employ a random permutation over the order in which the parts are generated.
Moreover, a configuration describing how the semi-definite programming relaxation encodes an edge involves $7$ different vectors (see Sections \ref{sec:sdp} and  \ref{sec:algorithm}).
Thus, we had to incorporate many technical ingredients, {\em e.g.}, analytically bounding the gradient of the separation probability and restricting the search to specific type of configurations while analytically bounding the error this incurs, to make the computer assisted proof terminate faster.
This results in about $ 150,000$ hours of CPU, which is roughly $20$ CPU years, to prove Theorem \ref{theorem:final}. 

\subsection{Additional Related Work}

The \maxbis problem has a long and rich history.
Frieze and Jerrum \cite{frieze1997improved} presented an approximation of $0.6514$ based on rounding a semi-definite program.
Later on, Ye \cite{Ye2001}, Halperin and Zwick \cite{halperin2002unified}, and Feige and Langberg \cite{feige2006rpr2} further improved the approximation guarantee to $0.699$, $0.7016$, and $0.7027$, respectively.
They achieved the above by strengthening the semi-definite programming relaxation, {\em e.g.}, by adding triangle inequality constraints, and presenting better rounding methods.
The next leap in approximating \maxbis came with the work of Raghavendra and Tan \cite{RaghavendraT12}.
They utilized a higher-level Lasserre hierarchy semi-definite program, together with an elegant rounding algorithm, and obtained a $0.85$-approximation.
Later on, Austrin, Benabbas, and Georgiou \cite{AustrinBG16} showed an improved rounding algorithm, pushing the approximation guarantee up to $0.877$.
It should be noted that the latter is very close to the best possibloe approximation of  $0.878$ for \maxcut.


Focusing on \maxkcut, Frieze and Jerrum \cite{frieze1997improved} presented an approximation algorithm with better guarantee than the naive random algorithm.
They utilized a semi-definite program relaxation alongside an elegant rounding algorithm that samples $k$ random vectors and assigns every vertex $v\in V$ to the cluster of the random vector that is closest to $v$'s vector in the relaxation.
Goemans and Williamson \cite{goemans2001approximation} presented an improved approximation of $ 0.836$ for \maxtricut by using a complex semi-definite program. In \cite{de2004approximate}, de Klerk, Pasechnik, and Warners presented further improved bounds for \maxkcut. Please refer to Table 1 by Newman \cite{Newman2018ComplexSP} for a summary of approximation guarantees.

\subsection{Paper Organization}
We start by presenting preliminary definitions in Section \ref{sec:prelim}. Next, in Section \ref{sec:sdp} we present our semi-definite program for \maxtris and show that it can be strengthened to obtain a solution which is globally uncorrelated. In Section \ref{sec:algorithm} we present our rounding algorithm and its analysis. To obtain a bound on the approximation guarantee of our rounding algorithm, we present an analysis which is based on a computer-assisted proof in Section \ref{Section:Experiment}. Furthermore, we discuss the generalization of our algorithm, and its numerical estimation, in Section \ref{sec:numerAndMaxkSec}.
Missing proofs appear in the appendix.

\section{Preliminaries}\label{sec:prelim}
{We denote by $\Phi:\mathbb{R}\rightarrow [0,1]$ the cumulative distribution function of the normal gaussian distribution and by $ \Phi ^{-1}:[0,1]\rightarrow \mathbb{R}$ its inverse.
Specifically, if $ R\sim N[0,1]$ then: $(1)$ $ \forall x\in \mathbb{R}$: $ \Pr[R\leq x]=\Phi(x)$ (or equivalently $ \Pr[R\geq x]=1-\Phi(x)$); and $(2)$ $\forall x\in [0,1] $: $ \Pr[R\leq \Phi^{-1}(x)]=x$ (or equivalently $ \Pr[R\geq \Phi^{-1}(x)]=1-x$).
Moreover, we say that a vector $\bg$ is a {\em random gaussian vector} if its coordinates are i.i.d standard gaussian $ N(0,1)$ random variables.

We denote by $ \Gamma _t:[0,1]^2\rightarrow [0,1]$ the probability that a standard bi-variate gaussian distribution with correlation $t$ has both its coordinates at most the given quantiles, {\em i.e.},
\[
\forall q_1,q_2\in [0,1]:~\Gamma _t (q_1,q_2)\triangleq \Pr [X\leq \Phi^{-1}(q_1),Y\leq \Phi^{-1}(q_2)] ,
\begin{pmatrix}
X \\
Y
\end{pmatrix}\sim N\left(\begin{pmatrix}
0 \\
0
\end{pmatrix},\begin{pmatrix}
1 & t \\
t & 1
\end{pmatrix}\right).
\]



We define the mutual information between two random variables.
\begin{definition}
    Let $X, Y$ be jointly distributed random variables taking values in $[q]$. The mutual information of $X$ and $Y$ is defined as 
    \[ I(X, Y) \triangleq \sum_{i, j \in [q]}\Pr(X = i, Y = j) \log \left( \frac{\Pr(X = i, Y = j)}{\Pr(X = i)\Pr(X = j)} \right). \]
\end{definition}


For any two disjoint subsets of vertices $ A,B\subseteq V$, we denote by $ \delta(A,B)$ the collection of edges having one endpoint in $A$ and another endpoint in $B$.
Hence, $|\delta(A,B)| $ denotes the number of edges crossing between $A$ and $B$.\footnote{For simplicity of presentation, we assume from this point onward that the graph is unweighted.
All of our results apply to graphs equipped with non-negative edge weights, in which case one should substitute $ |\delta(A,B)|$ with the total weight of edges crossing between $A$ and $B$.}

\section{The SDP Relaxation for \texorpdfstring{\maxtris}{Max-3-Section}} \label{sec:sdp}
In this section, we present a semi-definite programming ($ \mathrm{SDP}$) formulation and prove that it is a relaxation for the \maxtris problem.
Similarly to previous works on \maxbis, {\em e.g.}, \cite{AustrinBG16,RaghavendraT12}, we strengthen this formulation and obtain additional properties that will be useful to our rounding algorithm.
We define the following SDP formulation for \maxtris:
\begin{align}
(\mathrm{SDP})~~~~~~~\text{maximize   }~~~ &\sum_{e=(u,v) \in E}\left(1 - \sum_{i = 1}^3 \by_u^i \cdot \by_v^i  \right) \label{eq:SDP} \\
\text{s.t.   }~~~& \| \by_{\emptyset} \|^2 = 1\label{SDP:unit_vector} \\
&\|\by_v^i\|^2 = \by_{\emptyset} \cdot \by_v^i  &\forall v \in V, \forall i =1,2,3 \label{SDP:marginal} \\
 &\| \by_v^1 \|^2 +\| \by_v^2 \|^2+\| \by_v^3 \|^2= 1 &\forall v\in V \label{SDP:distribution} \\
 &\by_v^i \cdot \by_v^j = 0 & \forall v \in V, i \neq j \label{SDP:orthogonal} \\
&\by_u^i\cdot \by_v^j \geq 0 & \forall u,v\in V, \forall i,j=1,2,3 \label{SDP:nonnegative} \\
 & \sum_{v\in V}\|\by_v^i\|^2 = n/3 &\forall i=1,2,3 \label{SDP:balance}
 \end{align}
We note that in the above and what follows, for every vector $\by$ a square norm of a vector $\|\by\|^2$ is with respect to the $\ell_2$ (Euclidean) norm and equals $\by\cdot \by$.
Next, we prove that the formulation is a relaxation for our problem.
{Intuitively, for every vertex $v\in V$ the formulation $\mathrm{SDP}$ assigns a distribution over the three clusters via the vectors $\by_v^1$, $\by_v^2$, and $\by_v^3$.
Specifically, $\by_{\emptyset}$ is a unit vectors (Constraint (\ref{SDP:unit_vector})) that denotes {\em true} whereas the zero vector (that does not appear explicitly in $ \mathrm{SDP}$) denotes {\em false}.
Each vector $\by_v^i$ indicates how much vertex $v$ is likely to be assigned to the $i$\textsuperscript{th} cluster by $\mathrm{SDP}$.
Hence, $ \|\by_v^i\| ^2$, or equivalently $
\by_v^i\cdot \by_{\emptyset} $ (see Constraint (\ref{SDP:marginal})), denotes the marginal probability of assigning vertex $v$ to the $ i$\textsuperscript{th} cluster by $ \mathrm{SDP}$.
For every vertex $v\in V$, the sum of these marginal probabilities needs to sum up to one (Constraint \ref{SDP:distribution}).
Since every vertex $v\in V$ can be assigned to a single cluster in any integral solution, $ \mathrm{SDP}$ enforces that the vectors $ \by_v^i$ and $ \by_v^j$ for $ i\neq j$ are orthogonal (Constraint (\ref{SDP:orthogonal})).
Intuitively, the joint probability $\mathrm{SDP}$ assigns for vertices $u$ and $v$ to belong to the $i$\textsuperscript{th} and $j$\textsuperscript{th} clusters, respectively, is non-negative (Constraint (\ref{SDP:nonnegative})).
Finally, since the three clusters are required to be of size $ n/3$ each Constraint (\ref{SDP:balance}) is added to $\mathrm{SDP}$.
When focusing on the objective of $\mathrm{SDP}$ (see (\ref{eq:SDP})), for every edge $ (u,v)\in E$ the inner product $\by_u^i\cdot \by_v^i $ intuitively indicates the joint probability of both $u$ and $v$ to be assigned to the $ i$\textsuperscript{th} cluster by $\mathrm{SDP}$.
Therefore, intuitively $ 1-\by_u^1\cdot \by_v^1-\by_u^2\cdot \by_v^2-\by_u^3\cdot \by_v^3$ indicates the likelihood of separating an edge $ (u,v)$ by $ \mathrm{SDP}$.
Thus, this is the objective of $\mathrm{SDP}$.}

The following lemma proves that $\mathrm{SDP}$ is a relaxation to the \maxtris problem, {\em i.e.}, the value of an optimal solution $\mathrm{OPT}_{\mathrm{SDP}}$ to $\mathrm{SDP}$ is an upper bound on the value of an integral optimal solution $\mathrm{OPT}$.

\begin{lemma}
\label{claim:relaxation}
Given an instance of \maxtris let $\mathrm{OPT}_{\mathrm{SDP}}$ be the value of an optimal solution of $\mathrm{SDP}$ (\ref{eq:SDP}) and $\mathrm{OPT}$ be the value of an optimal integral solution.
Then, $\mathrm{OPT}_{\mathrm{SDP}} \geq \mathrm{OPT}$.
\end{lemma}
\begin{proof}
    Let $\{C^*_1, C^*_2, C^*_3\}$ be an optimal solution for the given instance of \maxtris whose value is $\mathrm{OPT}$.
    Construct the following vector solution to $\mathrm{SDP}$.
    First, fix an arbitrary unit vector $\by_{\emptyset}$.
    Second, for every $v\in V$  define $\by_v^i$ to be the zero vector if $v\notin C^*_i$ and $\by_v^i = \by_\emptyset$ if $v\in C^*_i$. 
    One may notice that all the constraints hold for this solution and the value of the objective of $ \mathrm{SDP}$ equals the value of the optimal solution $\{ C^*_1,C^*_2,C^*_3\}$.
    Hence, $\mathrm{OPT}_{\mathrm{SDP}} \geq \mathrm{OPT}$. 
\end{proof}

For simplicity of presentation, we denote by $Y$ a solution to $\mathrm{SDP}$.
Thus, $Y$ consists of $ \{ \by_u^i\}_{u\in V,i=1,2,3}$ and $\by_{\emptyset}$.
A useful property that any feasible solution $Y$ to $\mathrm{SDP}$ satisfies is that $ \by_u^1+\by_u^2+\by_u^3$ always equals the vector $ \by_{\emptyset}$.
This is summarized in the following lemma.

\begin{lemma}\label{lem:SDP_sum_vectors}
Let $Y$ be a feasible solution to $\mathrm{SDP}$.
Then, for every vertex $ u\in V$: $ \by_u^1+\by_u^2+\by_u^3 = \by_{\emptyset}$.    
\end{lemma}

An immediate corollary of the above lemma is that for every pair of vertices $ u,v\in V$: $ \by_u^i\cdot (\by_v^1+\by_v^2+\by_v^3) =\|\by_u^i \|^2$ (via Constraint (\ref{SDP:marginal})).

\subsection{Globally Uncorrelated Solution}\label{Section:Uncorrelated}

Next we define when a solution $Y$ to $\mathrm{SDP}$ is globally uncorrelated following the framework of ~\cite{RaghavendraT12}. Global uncorrelation implies that our rounding algorithm returns a solution that has close to $\frac{n}{3}$ vertices in each part with high probability (see Lemma~\ref{lem:concentration}). The sizes then can be corrected with a minor loss in approximation ratio by randomly shifting the imbalanced vertices (see Lemma~\ref{lemma:balancing}).

A simple fact about any $\mathrm{SDP}$ solution is the following: given any two vertices $u,v$, there exists a \emph{local} probability distribution $\mu_{u,v}$ on $\{ 1,2,3\}^2$ such that $\Pr_{\mu_{u,v}}[X_u=i ,X_v=j]= \langle \by_u^i, \by_v^j\rangle$ for every $i,j\in \{ 1,2,3\}$ and $\Pr_{\mu_{u,v}}[X_u=i]=\|\by_u^i\|^2$ for every $i\in \{ 1,2,3\}$. The distribution implies that, at least, locally the semi-definite program is a distribution over integral solutions that satisfy correct correlations. The last property states that the distributions are consistent on their intersection which can be at most one vertex. 

\begin{definition}\label{def:independent}
A solution $Y$ to $\mathrm{SDP}$ is $\varepsilon$-independent if $\EE_{u,v}[I_{\mu_{u,v}}(X_u, X_v)]\leq \varepsilon$ where $\mu_{u,v}$ is the local probability distribution associated with vertices $u$ and $v$.  
\end{definition}

The following lemma is an application of Theorem 4.6 from \cite{RaghavendraT12} to {our $\mathrm{SDP}$ for \maxtris} and it shows how to obtain a $\varepsilon$-independent solution. The algorithm proving Lemma~\ref{lem:hierarchy} follows from solving the {$\Theta(t^2)$-level}
Lasserre hierarchy semi-definite program for $\mathrm{SDP}$ and then inductively conditioning on variables. 

\begin{lemma}\label{lem:hierarchy}
There is an algorithm which, given an integer $t>0$ and {an instance of \maxtris}, runs in
time $n^{O(poly(t))}$ and outputs a set of vectors $Y$ consisting of $\{\by_v^i\} _{v\in V,i=1,2,3}$ and $\by_{\emptyset}$  such that:
\begin{enumerate}
    \item $Y$ is a feasible solution to $\mathrm{SDP}$. 
    \item The objective value of $\mathrm{SDP}$ (\ref{eq:SDP}) when plugging in $Y$ is at least $ {\mathrm{OPT}_{\mathrm{SDP}}-  \frac{1}{t}}$.
    \item $Y$ is  $\frac1t$-independent.
\end{enumerate} 
\end{lemma}


\section{The Rounding Algorithm}\label{sec:algorithm}

In this section we present our rounding algorithm for $\mathrm{SDP}$, which appears in Algorithm \ref{Algorithm:Max3SecAlg}.
The algorithm receives as input a solution to $\mathrm{SDP}$ and outputs a partition $\{ C_1,C_2,C_3\}$ of $V$ that: $(1)$ has high value compared to the $\mathrm{SDP}$ value of the input solution; and $(2)$ is (nearly) balanced, {\em i.e.}, for every $ i=1,2,3$: $ |C_i|$ is close to $ n/3$.
We will conclude the analysis by proving that one can re-balance the partition without a significant loss in the value of the solution.

In order to state the algorithm, we require the following definition.
For every vertex $u\in V$ and $i=1,2,3$ we denote by $\bz_u^i$ the normalized component of $ \by_u^i$ that is orthogonal to $ \by_{\emptyset}$, {\em i.e.}, $ \by_u^i = \| \by_u^i\|^2 \by_{\emptyset} + \sqrt{\| \by_i^i\|^2 - \| \by_u^i\|^4} \bz_u^i$.
Equivalently,
\begin{align}
     \bz_u^i \triangleq \frac{\by_u^i - \| \by_u^i\|^2 \by_{\emptyset}}{\sqrt{\| \by_u^i\|^2 - \| \by_u^i\|^4}} .\label{def_z}
\end{align}
Clearly, $ \bz_u^i$ is a unit vector that is orthogonal to $ \by_{\emptyset}$.
We note that if the marginal of vertex $u$ and cluster $i$ is integral, {\em i.e.}, $ \| \by_u^i\|^2\in \{ 0,1\}$, then $ \bz_u^i$ is not defined.
In this case one can simply choose an arbitrary unit vector in the space orthogonal to $ \by _{\emptyset}$ to be $ \bz_u^i$.

\SetKwInput{KwInput}{Input}                
\SetKwInput{KwOutput}{Output}              

\begin{algorithm}[H]
\caption{Max-3-Section Rounding Algorithm}
\DontPrintSemicolon
\label{Algorithm:Max3SecAlg}
\KwInput{solution $\{\by_u^i\} _{u\in V,i=1,2,3}$ and $\by_{\emptyset} $ to $ \mathrm{SDP}$.}
\KwOutput{a partition of $V$ into three parts.}

Draw uniformly at random a permutation $\pi \in \mathcal{S}_3$.\;

Draw independently two random Gaussian vectors $\bg_1$ and $ \bg_2 $.\; 

Define the following sets:
\begin{align*}
    S_{\pi(1)} &\triangleq \left\{ u \in V : \bz_u^{\pi(1)} \cdot \bg_1 \geq \Phi^{-1} \left(1-\|\by_u^{\pi(1)}\|^2\right) \right\}, \\
S_{\pi(2)} &\triangleq \left\{ u \in V : \bz_u^{\pi(2)} \cdot \bg_2 \geq \Phi^{-1}\left(1- \|\by_u^{\pi(2)}\|^2 / (1-\|\by_u^{\pi(1)}\|^2) \right) \right\} .
\end{align*}

Return $ \{ C_1,C_2,C_3\}$ where: $C_{\pi(1)}\triangleq S_{\pi(1)}$, $C_{\pi(2)} \triangleq  S_{\pi(2)} \setminus S_{\pi(1)}$, $C_{\pi(3)} \triangleq  V\setminus (S_{\pi(1)} \cup S_{\pi(2)})$.

\end{algorithm}



We first prove that Algorithm \ref{Algorithm:Max3SecAlg} preserves the marginal probabilities of $\mathrm{SDP}$, {\em i.e.}, $\| \by_u^i\|^2$ is the probability vertex $u$ is assigned to cluster $ C_i$.
This is summarized in the following lemma.
\begin{lemma}
\label{lem:marginals}
For every $u\in V$ and $i=1,2,3$, it holds that $\Pr[u \in C_i] = \|\by_u^i\|^2.$
\end{lemma}

\begin{proof}
Fix a permutation $\pi \in \mathcal{S}_3$, and let us calculate the probability that $u \in C_i$ conditioned on the event that $\pi$ was chosen in the first step of Algorithm \ref{Algorithm:Max3SecAlg}.
Hence, the following holds for $ C_{\pi(1)}$:
\[ \Pr \left[u \in C_{\pi(1)} | \pi\right] = \Pr\left[u \in S_{\pi(1)} | \pi\right] = \Pr \left[\bz_u^{\pi(1)}\cdot \bg_1\geq \left(1-\Phi^{-1}\left(1-\|\by_u^{\pi(1)}\|^2\right)\right)|\pi\right]=\| \by_u^{\pi(1)}\|^2.\]
We observe that the sets $S_{\pi(1)}$ and $S_{\pi(2)}$ are constructed with independent vectors $\bg_1$ and $\bg_2$. 
Therefore, similarly to the above, the following holds for $ C_{\pi(2)}$:
\begin{align*}
\Pr\left[u \in C_{\pi(2)} | \pi\right] &= \Pr\left[u \notin S_{\pi(1)} \wedge u \in S_{\pi(2)} | \pi\right] \\
&= \Pr \left[ u \notin S_{\pi(1)}|\pi\right] \cdot \Pr \left[ u \in S_{\pi(2)}|\pi\right]\\
& = \left( 1-\|\by_u^{\pi(1)}\|^2\right) \cdot \Pr \left[\bz_u^{\pi(2)}\cdot \bg_2\geq \left(1-\Phi^{-1}\left(1-\frac{\|\by_u^{\pi(2)}\|^2}{(1-\| \by_u^{\pi(1)}\|^2)}\right)\right)\Big|\pi\right]\\
& = \left( 1-\|\by_u^{\pi(1)}\|^2\right) \cdot \frac{\|\by_u^{\pi(2)}\|^2}{(1-\| \by_u^{\pi(1)}\|^2)} \\
& = \| \by_u^{\pi(2)}\|^2 .
\end{align*}
Finally, since the events $ \{u\in C_{\pi(1)}|\pi\}$, $ \{u\in C_{\pi(2)}|\pi\}$, and $ \{ u\in C_{\pi(3)}|\pi\}$ are disjoint and exactly one of them occurs, {\em i.e.}, every vertex $ u\in V$ belongs to exactly one cluster in the output, we can conclude that:
\[ \Pr\left[u\in C_{\pi(3)}|\pi\right]=1-\Pr\left[u\in C_{\pi(1)}|\pi\right]-\Pr\left[u\in C_{\pi(2)}|\pi\right]=1-\| \by_u^{\pi(1)}\| ^2- \| \by_u^{\pi(2)}\|^2=\| \by_u^{\pi(3)}\|^2.\]
In the above the last equality follows from Constraint (\ref{SDP:distribution}).
Unfixing the conditioning on $\pi$ by using the law of total probability concludes the proof.
\end{proof}

Our goal is to write an expression for the probability that an edge crosses between two different parts in the partition that Algorithm \ref{Algorithm:Max3SecAlg} outputs.
Given a fixed pair of vertices $ u,v\in V$ and $ i=1,2,3$, we denote by $t_i$ the inner product between $ \bz _u^i$ and $ \bz_v^i$.
One should note that Constraint \ref{SDP:marginal} in SDP \ref{eq:SDP} implies:
\begin{align} t_i&=  \frac{(\by_u^i-x_i\cdot \by_\emptyset)\cdot(\by_v^i-w_i\cdot \by_\emptyset)}{\sqrt{(x_i-x_i^2)\cdot(w_i-w_i^2)}} = \frac{\alpha_i -x_iw_i}{\sqrt{(x_i-x_i^2)\cdot(w_i-w_i^2)}},\label{def_t}
\end{align}
where $ x_i\triangleq \| \by_u^i\|^2$ and $ w_i\triangleq \| \by_v^i\|^2$ are the marginal values the $\mathrm{SDP}$ assigns to vertices $u$ and $v$, respectively, with respect to the $i$\textsuperscript{th} cluster, and $ \alpha _i\triangleq \by _u^i \cdot \by_v^i$ is the correlation the $\mathrm{SDP}$ assigns for both $u$ and $v$ being assigned to the $i$\textsuperscript{th} cluster.
The following lemma gives the desired expression.
We require the following claim for its proof:
\begin{claim}\label{claim:bivariate_gaussian_gamma}
Let $(X,Y)$ be a standard bi-variate Gaussian with correlation $t$.
Then for every $q_1, q_2 \in [0, 1]$, we have
$ \Pr[X \geq \Phi^{-1}(1 - q_1), Y \geq \Phi^{-1}(1 - q_2)] = \Gamma_t(q_1, q_2).  $
\end{claim}

\begin{lemma} \label{lem:cutProb}
For every $ u,v\in V$, let $\mathcal{A}_{u,v}$ be the event that Algorithm \ref{Algorithm:Max3SecAlg} separates $u$ and $v$:

\begin{align*}
 \Pr\left[\mathcal{A}_{u,v}\right] = & 1 - \frac{1}{6} \sum_{ \pi \in \mathcal{S}_3} \Bigg[ \Gamma_{t_{\pi(1)}}\left(x_{\pi(1)}, w_{\pi(1)}\right) + \Gamma_{t_{\pi(1)}}\left( 1- x_{\pi(1)},1 - w_{\pi(1)}\right)\cdot\\ 
   & \cdot \left(  \Gamma_{t_{\pi(2)}}\left(1 - \frac{x_{\pi(2)}}{1 - x_{\pi(1)}}, 1 - \frac{w_{\pi(2)}}{1 - w_{\pi(1)}}\right) + \Gamma_{t_{\pi(2)}}\left(\frac{x_{\pi(2)}}{1 - x_{\pi(1)}}, \frac{w_{\pi(2)}}{1 - w_{\pi(1)}}\right)  \right)  \Bigg].     
\end{align*}
\end{lemma}

\begin{proof}[Proof of Lemma \ref{lem:cutProb}] 
Fix a permutation $\pi$.
For this fixed permutation $\pi$, vertices $u$ and $v$ are {\em not} separated by Algorithm \ref{Algorithm:Max3SecAlg} if both vertices are in $ C_{\pi(1)}$, or $ C_{\pi(2)}$, or $ C_{\pi(3)}$.
Let us write the probability of these three events.
First,
\begin{align*}
\Pr\left[u \in C_{\pi(1)}, v \in C_{\pi(1)}|\pi\right] & = \Pr\left[u \in S_{\pi(1)}, v \in S_{\pi(1)}|\pi\right] \\
&=\Pr\left[\bg_1\cdot  \bz_u^{\pi(1)} \geq \Phi^{-1}\left(1 - x_{\pi(1)}\right), \bg_1\cdot \bz_v^{\pi(1)} \geq \Phi^{-1}\left(1 - w_{\pi(1)}\right) |\pi\right]\\
&= \Gamma_{t_{\pi(1)}}\left(x_{\pi(1)}, w_{\pi(1)}\right) .
\end{align*}
The last equality follows from Claim \ref{claim:bivariate_gaussian_gamma}, since $ (\bg_1\cdot  \bz_u^{\pi(1)},\bg_1\cdot \bz_v^{\pi(1)})$ is a standard bi-variate gaussian with correlation $ t_{\pi(1)}$.

Second, both $u$ and $v$ are in $ C_{\pi(2)}$ if both are not in $ S_{\pi(1)}$ and both are in $ S_{\pi(2)}$.
The former and latter events are independent since $ \bg_1$ and $ \bg_2$ are chosen independently in Algorithm \ref{Algorithm:Max3SecAlg}.
Hence,
\begin{align*}
\Pr \left[ u\in C_{\pi(2)},v\in C_{\pi(2)}|\pi\right] & = \Pr \left[ u\notin S_{\pi(1)},v\notin S_{\pi(1)},u\in S_{\pi(2)}, v\in S_{\pi(2)}|\pi\right] \\
& = \Pr \left[ u\notin S_{\pi(1)},v\notin S_{\pi(1)}|\pi\right] \cdot \Pr \left[ u\in S_{\pi(2)}, v\in S_{\pi(2)}|\pi\right] .
\end{align*}
The probability of the former event equals:
\begin{align*}
\Pr\left[u \notin S_{\pi(1)}, v \notin S_{\pi(1)}|\pi\right] & =  \Pr\left [\bg_1\cdot \bz_u^{\pi(1)} \leq \Phi^{-1}\left(1 - x_{\pi(1)}\right), \bg_1\cdot \bz_v^{\pi(1)} \leq \Phi^{-1}\left(1 - w_{\pi(1)}\right)|\pi \right] \\
& = \Gamma_{t_{\pi(1)}}\left(1- x_{\pi(1)}, 1 - w_{\pi(1)}\right),
\end{align*}
whereas the probability of the latter event equals:
\begin{align*}
\Pr \left[ u\in S_{\pi(2)}, v\in S_{\pi(2)}|\pi\right] & = \Pr \left[ \bg_2 \cdot \bz_u^{\pi(2)}\geq \Phi^{-1}\left( 1-\frac{x_{\pi(2)}}{1-x_{\pi(1)}}\right), \bg_2\cdot \bz_v^{\pi(2)}\geq \Phi^{-1}\left( 1-\frac{w_{\pi(2)}}{1-w_{\pi(1)}}\right)|\pi\right]  \\
& =  \Gamma_{t_{\pi(2)}}\left(\frac{x_{\pi(2)}}{1 - x_{\pi(1)}}, \frac{w_{\pi(2)}}{1 - w_{\pi(1)}} \right).
\end{align*}
As before, last equality follows from Claim \ref{claim:bivariate_gaussian_gamma}, since $ (\bg_2\cdot  \bz_u^{\pi(2)},\bg_2\cdot \bz_v^{\pi(2)})$ is a standard bi-variate gaussian with correltation $ t_{\pi(2)}$.
Thus, we can conclude that:
$$ \Pr \left[ u\in C_{\pi(2)},v\in C_{\pi(2)}|\pi\right] = \Gamma_{t_{\pi(1)}}\left(1- x_{\pi(1)}, 1 - w_{\pi(1)}\right) \cdot \Gamma_{t_{\pi(2)}}\left(\frac{x_{\pi(2)}}{1 - x_{\pi(1)}}, \frac{w_{\pi(2)}}{1 - w_{\pi(1)}} \right).$$

Third, both $u$ and $v$ are in $ C_{\pi(3)}$ if both are not in $ S_{\pi(1)}$ and both are not in $S_{\pi(2)} $.
The former and latter events are independent since $\bg_1$ and $\bg_2$ are chosen independently in Algorithm \ref{Algorithm:Max3SecAlg}.
Hence,
\begin{align*}
\Pr \left[ u\in C_{\pi(3)},v\in C_{\pi(3)} |\pi\right] & = \Pr \left[ u\notin S_{\pi(1)},v\notin S_{\pi(1)},u\notin S_{\pi(2)},v\notin S_{\pi(2)}|\pi\right] \\& = \Pr \left[ u\notin S_{\pi(1)},v\notin S_{\pi(1)} |\pi\right]\cdot \Pr \left[ u\notin S_{\pi(2)},v\notin S_{\pi(2)}|\pi\right] .  
\end{align*}
Recall, as proved above, that the probability of the former event equals:
\begin{align*}
\Pr \left[ u\notin S_{\pi(1)},v\notin S_{\pi(1)}|\pi\right] & =  \Gamma_{t_{\pi(1)}}\left(1- x_{\pi(1)}, 1 - w_{\pi(1)}\right), 
\end{align*}
whereas the probability of the latter event equals:
\begin{align*}
\Pr \left[ u\notin S_{\pi(2)}, v\notin S_{\pi(2)}|\pi\right] & = \Pr \left[ \bg_2 \cdot \bz_u^{\pi(2)}\leq \Phi^{-1}\left( 1-\frac{x_{\pi(2)}}{1-x_{\pi(1)}}\right), \bg_2\cdot \bz_v^{\pi(2)}\leq \Phi^{-1}\left( 1-\frac{w_{\pi(2)}}{1-w_{\pi(1)}}\right)|\pi\right]  \\
& =  \Gamma_{t_{\pi(2)}}\left(1-\frac{x_{\pi(2)}}{1 - x_{\pi(1)}}, 1-\frac{w_{\pi(2)}}{1 - w_{\pi(1)}} \right).
\end{align*}
Thus, we can conclude that:
$$\Pr\left[u \in C_{\pi(3)}, v \in C_{\pi(3)}|\pi\right] = \Gamma_{t_{\pi(1)}}\left(1- x_{\pi(1)}, 1 - w_{\pi(1)}\right) \cdot \Gamma_{t_{\pi(2)}}\left(1 -\frac{x_{\pi(2)}}{1 - x_{\pi(1)}}, 1 -  \frac{w_{\pi(2)}}{1 - w_{\pi(1)}}\right).$$
The proof is concluded since: $$ \Pr \left[ \text{Algorithm \ref{Algorithm:Max3SecAlg} separates $u$ and $v$}\right] = 1-\frac{1}{6}\sum _{\pi \in \mathcal{S}_3}\sum _{j=1}^3 \Pr \left[ u\in C_{\pi(j)}, v\in C_{\pi(j)}|\pi\right] .$$ 

\end{proof}

Our goal now is to lower bound the expected value of the output of Algorithm \ref{Algorithm:Max3SecAlg}, before it is re-balanced (with a negligible loss) to ensure the size of each cluster is exactly $ n/3$.
As our analysis is performed edge-by-edge, {\em i.e.}, for every edge we lower bound the ratio of the probability Algorithm \ref{Algorithm:Max3SecAlg} separates the edge to the contribution of this edge to the objective of $\mathrm{SDP}$ (\ref{eq:SDP}), we introduce the notions of a configuration and a feasible configuration.


A \emph{configuration} is a vector $\bc=(x_1, x_2, x_3, w_1, w_2, w_3, \alpha_1, \alpha_2, \alpha_3, t_1, t_2, t_3) \in \RR^{12}$, such that for every $ i=1,2,3$: $x_i, w_i, \alpha_i \in [0,1]$ and $t_i \in [-1,1]$. We say that a configuration $\bc$ is a \emph{feasible configuration} if it can be realized by vectors in a feasible solution to $\mathrm{SDP}$ (as the following definition states).
\begin{definition}\label{definition:feasible_config}
A configuration $ \bc=(x_1, x_2, x_3, w_1, w_2, w_3, \alpha_1, \alpha_2, \alpha_3, t_1, t_2, t_3)\in [0,1]^9 \times [-1,1]^3$ is called a {\em feasible configuration} if there are vectors $\by_u^1$, $\by_u^2$, $\by_u^3$, $\by_v^1$, $\by_v^2$, $\by_v^3$ and $ \by_{\emptyset}$ satisfying:
\begin{enumerate}
    \item The vectors $\by_u^1$, $\by_u^2$, $\by_u^3$, $\by_v^1$, $\by_v^2$, $\by_v^3$ and $ \by_{\emptyset}$ satisfy Constraints (\ref{SDP:unit_vector}) to (\ref{SDP:nonnegative}) in $ \mathrm{SDP}$.
    \item $x_i=\|\by_u^{i}\|^2$, $w_i=\|\by_v^{i}\|^2$ and $\alpha_i =\by_u^i \cdot \by_v^i$, $\forall i=1,2,3$.
    \item $t_i = (\alpha_i - x_i w_i)/((x_i - x_i^2)(w_i - w_i^2))^{1/2}$, $\forall i=1,2,3$.
\end{enumerate}
The vectors $\by_u^1$, $\by_u^2$, $\by_u^3$, $\by_v^1$, $\by_v^2$, $\by_v^3$ and $ \by_{\emptyset}$ are called a {\em realization} of $\bc$.
The set of all feasible configuration is denoted by $\C$.
\end{definition} 
We note that there is some redundancy in the above definition.
First, we can reduce the dimension of the configuration by two simply by substituting $x_3$  with $1-x_1-x_2$ and $w_3$ with $ 1-w_1-w_2$.
Second, we can remove $t_1$, $t_2$ and $t_3$ (or $\alpha_1$, $\alpha_2$ and $\alpha _3$) since each $ t_i$ (or $ \alpha _i$) can be derived from all the parameters excluding $ t_1$, $t_2$, and $ t_3$ (or excluding $\alpha _1$, $ \alpha _2$, and $ \alpha _3$). In Lemma~\ref{lem:feasibleConfigurations}, we give a characterization of $\C$  after projecting out  $t_1,t_2,t_3$. This description consists of linear constraints and we utilize the description for our computer-assisted proof. 
For simplicity of the analysis, we will keep all the parameters.

Let us now define two functions over $\C$.
The first function $f$, given a feasible configuration $ \bc\in \C$, returns the probability that Algorithm \ref{Algorithm:Max3SecAlg} cuts an edge whose associated vectors are a realization of $ \bc$.
Formally, following Lemma \ref{lem:cutProb}:
\begin{align*}
 f(\bc) \triangleq & 1 - \frac{1}{6} \sum_{ \pi \in \mathcal{S}_3} \Bigg[ \Gamma_{t_{\pi(1)}}\left(x_{\pi(1)}, w_{\pi(1)}\right)  + \Gamma_{t_{\pi(1)}}\left( 1- x_{ \pi(1)},1 - w_{\pi(1)}\right) \cdot\\
 & \cdot \left(  \Gamma_{t_{\pi(2)}}\left(1 - \frac{x_{\pi(2)}}{1 - x_{\pi(1)}}, 1 - \frac{w_{\pi(2)}}{1 - w_{\pi(1)}}\right) + \Gamma_{t_{\pi(2)}}\left(\frac{x_{\pi(2)}}{1 - x_{\pi(1)}}, \frac{w_{\pi(2)}}{1 - w_{\pi(1)}}\right)  \right)              \Bigg].    
\end{align*}
The second function $g$, given a feasible configuration $\bc \in \C$, returns the contribution to the objective of $\mathrm{SDP}$ of an edge whose associated vectors are a realization of $ \bc$.
Formally, following (\ref{eq:SDP}):
\[ g(\bc) \triangleq 1-\alpha_1 - \alpha _2 - \alpha _3.\]
It is important to note that both $f$ and $g$ can be evaluated for every configuration $\bc \in [0,1]^9 \times [-1,1]^3$ which might not be necessarily feasible.
However, for such a (non feasible) configuration $\bc$, $ f(\bc)$ and $ g(\bc)$ lose their ``meaning''.




{
To lower bound the value of the solution $\{ C_1, C_2, C_3\}$ Algorithm \ref{Algorithm:Max3SecAlg} outputs, we introduce the following:
\begin{align}
    \mu \triangleq \inf _{\bc \in \C} \left\{ \frac{f(\bc)}{g(\bc)}\right\} .\label{approx_mu}
\end{align}
Clearly, from the above definition of $\mu$, the value of the output $\{ C_1, C_2, C_3\}$ of Algorithm \ref{Algorithm:Max3SecAlg} is at least: $ \mu \cdot \mathrm{OPT}_{\mathrm{SDP}}\geq \mu \cdot \mathrm{OPT}$ (where the inequality follows from Lemma \ref{claim:relaxation} which states that $\mathrm{SDP}$ is a relaxation).
Hence, if the output  $\{ C_1, C_2, C_3\}$ of Algorithm \ref{Algorithm:Max3SecAlg} was perfectly balanced, {\em i.e.}, $|C_1|=|C_2|=|C_3|=n/3$, Algorithm \ref{Algorithm:Max3SecAlg} would achieve an approximation of $\mu$ to the \maxtris problem.
In what follows we show that one can re-balance $\{ C_1, C_2, C_3\}$ without a significant loss in the value of the solution.
Thus, our goal is to lower bound $\mu$.
For now, as formally proving the exact value of $\mu$ is a challenging task, we state the following conjecture that follows from numeric estimation of $\mu$.
\begin{conjecture}\label{conj:mu}
    $\mu\geq 0.8192$.
\end{conjecture}
}

Assuming the above conjecture regarding $\mu$ (Conjecture \ref{conj:mu}), there are two things that are left in order to conclude the analysis of our algorithm.
First, we show that if the solution $Y$ to $\mathrm{SDP}$ is independent (as in Definition \ref{def:independent} and Lemma \ref{lem:hierarchy} ) 
then with a sufficiently high probability every cluster $ C_i$ is close to the desired size of $ n/3$. 
This gives rise to the following Definition \ref{def:unblanaced} and Lemma \ref{lem:concentration}.
Second, we show that a solution $\{ C_1,C_2,C_3\}$ that is close to being perfectly balanced can be efficiently re-balanced without a significant loss in its value.
The latter is summarized in Lemma \ref{lemma:balancing}.



\begin{definition}\label{def:unblanaced}
    A partition $\{C_1, C_2, C_3\}$ of a graph on $n$ nodes is {\em $\varepsilon$-unbalanced} if for every $i=1,2,3$:
    \[\frac{n}{3}(1 - \varepsilon) \leq |C_i| \leq \frac{n}{3}(1 + \varepsilon).\]
\end{definition}


\begin{lemma}\label{lem:concentration}
Let $Y$ be a $\frac{1}{t}$-independent solution to $\mathrm{SDP}$ where $ t=\Omega(\varepsilon ^{-18})$, and $\{C_1,C_2,C_3\}$ be the partition that Algorithm \ref{Algorithm:Max3SecAlg} outputs on $Y$. Then for every $i=1,2,3$ it holds that:
\[Pr\left[\left||C_i|-n/3\right|\geq \frac{\varepsilon n}{3} \right] \leq \varepsilon.\]
\end{lemma}

Next, we show that such unbalanced partition can be balanced without a large loss in the objective. That is, we present an algorithm that given a $\epsilon$-unbalanced partition, finds in polynomial time a balanced partition with small loss in the objective, in expectation.

\begin{lemma} \label{lemma:balancing}
There is a polynomial-time algorithm that given a $\epsilon$-unbalanced partition $\{C_1,C_2,C_3\}$ with value $\Delta = |\delta(C_1, C_2)| + |\delta(C_2, C_3)| + |\delta(C_1, C_3)|$ finds a balanced partition $\{C_1', C_2', C_3'\}$ with expected value $\mathbb{E}[\Delta'] \geq (1 - 2\epsilon)\Delta$.
\end{lemma}

The proofs of Lemma \ref{lem:concentration} and Lemma \ref{lemma:balancing} appear in Appendix \ref{apx:proofsSec4}. We combine these lemmas and prove the following result.

\begin{theorem}\label{thrm:mu_approximation}
    For every constant $\varepsilon > 0$, there exists a polynomial-time approximation algorithm for \maxtris, that runs in time $ n^{O(\text{poly}(\varepsilon ^{-1}))}$, achieving an approximation of {$(1-2\epsilon)(\mu-O(\varepsilon))$}.
\end{theorem}

{
\begin{proof}
    Let $\varepsilon >0$ be a constant, and $t$ an integer satisfying $t=\Omega(\varepsilon ^{-18})$.
Lemma \ref{lem:hierarchy} shows that we can compute in polynomial time a solution $Y$ to $\mathrm{SDP}$  that is $\frac{1}{t}$-independent with only an additive loss of $ 1/t$ in the objective.
We repeatedly apply Algorithm \ref{Algorithm:Max3SecAlg} to round the above $Y$ until we obtain a solution $\{ C_1,C_2,C_3\}$ that is $\varepsilon$-unbalanced, and then apply Lemma \ref{lemma:balancing} to re-balance it an obtain our final output.

Let us now analyze the approximation guarantee of the above algorithm.
First, It follows form Lemma \ref{lem:concentration}
and a simple union bound over the three clusters that with a probability of at least $ 1-3\varepsilon$, applying Algorithm \ref{Algorithm:Max3SecAlg} to round the above solution $Y$ yields a clustering $\{ C_1,C_2,C_3\}$ that is $\varepsilon$-unbalanced (as in Definition \ref{def:unblanaced}).
Let us denote by $\mathcal{A}$ the event that $ \{ C_1,C_2,C_3\}$ is $\varepsilon$-unbalanced.
Hence, $\Pr[\mathcal{A}]\geq 1-3\varepsilon$ and $\Pr[\bar{\mathcal{A}}] \leq 3\varepsilon$.
Moreover, as before, let us denote by $\Delta$ the value of the solution $ \{ C_1,C_2,C_3\}$.
Thus, the expected value of this solution {\em conditioned} on it being $\varepsilon$-unbalanced is at least:
\[
    \EE[\Delta|\mathcal{A}] = \frac{\EE[\Delta]-\Pr[\bar{\mathcal{A}}]\cdot \EE[\Delta|\bar{\mathcal{A}}]}{\Pr[\mathcal{A}]} \geq \mu \cdot  \mathrm{OPT}_{\mathrm{SDP-H}}-\frac{9\varepsilon}{2}\mathrm{OPT}.
\]
The inequality follows from the facts that: $(1)$ $\mathbb{E}[\Delta|\bar{\mathcal{A}}]\leq (3/2)\cdot \mathrm{OPT}$ (since $\Delta \leq m$ and $ \mathrm{OPT}\geq (2/3)m$); $(2)$ $ \mathbb{E}[\Delta]\geq \mu \cdot \mathrm{OPT}_{\mathrm{SDP-H}}$ (definition of $\mu$ (\ref{approx_mu})); and $(3)$ $ \Pr[\mathcal{A}]\leq 1, \Pr[\bar{\mathcal{A}}]\leq 3\varepsilon$.
We note that $\mathrm{OPT}_{\mathrm{SDP-H}} \geq \mathrm{OPT}_{\mathrm{SDP}}-\varepsilon^{18} \geq \mathrm{OPT} -\varepsilon^{18}$.
Hence,
\[\EE[\Delta|\mathcal{A}] \geq \mathrm{OPT}(\mu - O(\varepsilon)).\]

Applying Lemma \ref{lemma:balancing} concludes the proof.
\end{proof} }

Moreover, in Observation \ref{obs:worstConf} in the appendix we present a configuration $\bc$ that has a ratio of $\frac{f(\bc)}{g(\bc)}=0.8192$, hence that is an upper bound on $\mu$. 

\subsection{Towards Estimating \texorpdfstring{$\mu$}{u} via a Computer Assisted Proof}

For our computer assisted proof, we consider a slightly different version of $\mu$ which speeds up our code. Consider the following, for some fixed $\delta' > 0$:
\begin{align}
    \mu' \triangleq \inf _{\bc \in \C, g(\bc) \geq \delta'} \left\{ \frac{f(\bc)}{g(\bc)}\right\} .\label{approx_mu'}
\end{align}
The following lemma shows the loss we incur when using $\mu'$ rather than $\mu$ is bounded.
\begin{lemma} \label{lemma:approx-final}
    Let $\{C_1, C_2, C_3\}$ be the output of Algorithm \ref{Algorithm:Max3SecAlg} when run on a solution $Y$ to $\mathrm{SDP}$ \ref{eq:SDP} with objective value $\mathrm{SDP}_{\mathrm{VAL}}$. 
    Then we have that 
    \[ \EE[|\delta(C_1, C_2)| + |\delta(C_1, C_3)| + |\delta(C_2, C_3)|]  \geq \left(1 - \frac{\delta'}{2(1 - \delta')}\right) \mu' \cdot\mathrm{SDP}_{\mathrm{VAL}}.   \]
\end{lemma}

{
The following theorem summarizes the approximation guarantee when $\mu'$ is used instead of $\mu$. 
}

\begin{theorem} \label{theorem:finalAlg}
For any {constant} $\varepsilon > 0$, there is an algorithm that outputs a partition of the vertex set $\{C_1, C_2, C_3\}$ with $|C_1| = |C_2| = |C_3|$ satisfying, 
 \[ \mathbb{E}[|\delta(C_1, C_2)| + |\delta(C_1, C_3)| + |\delta(C_2, C_3)| ] \geq (1 - 2\epsilon)\left(1 - \frac{\delta'}{2(1 - \delta')}\right) \mu'(\mathrm{OPT}_{\mathrm{SDP}} - O(\epsilon^{3/2})) \] 
 with an expected run-time of $n^{O(1/\varepsilon)}$.
\end{theorem}
\begin{proof}
    First we use Lemma \ref{lem:hierarchy} to get a $\frac{1}{t}$-independent solution $Y$ for $t = \Omega(\epsilon^{-18})$. Let  $\mathrm{OPT}_{\mathrm{SDP-H}}$ 
    denote the objective value of this solution. Next we run Algorithm \ref{Algorithm:Max3SecAlg} on $Y$ and repeat until it outputs sets $C_1', C_2', C_3'$ that are $\epsilon$-unbalanced.    We note that by Lemma \ref{lem:concentration} $(C_1', C_2', C_3')$ is \emph{not} $\epsilon$-unbalanced with probability at most $3\epsilon$, so we run Algorithm \ref{Algorithm:Max3SecAlg} at most $1/3\epsilon$ times in expectation. Since $C_1', C_2', C_3'$ are $\epsilon$-unbalanced, applying the random shifting of Lemma \ref{lemma:balancing} to $C_1', C_2', C_3'$ we get $C_1, C_2, C_3$ satisfying 
    \[ \mathbb{E}[|\delta(C_1, C_2)| + |\delta(C_1, C_3)| + |\delta(C_2, C_3)| ] \geq (1 - 2 \epsilon)  \mathbb{E}[|\delta(C_1', C_2')| + |\delta(C_1', C_3')| + |\delta(C_2', C_3')|], \]
    From Lemma \ref{lemma:approx-final} we have, 
    \[\mathbb{E}[|\delta(C_1, C_2)| + |\delta(C_1, C_3)| + |\delta(C_2, C_3)| ] \geq (1 - 2\epsilon)\left(1 - \frac{\delta'}{2(1 - \delta')}\right)\mu'\mathrm{OPT}_{\mathrm{SDP-H}}. \] Then applying second item of Lemma \ref{lem:hierarchy} gives us that $\mathrm{OPT}_{\mathrm{SDP-H}} \geq \mathrm{OPT}_{\mathrm{SDP}} - O(\epsilon^{3/2})$ since we set $t = \Omega(\epsilon^{-18})$.  
\end{proof}


\section{Computer Assisted Proof} \label{Section:Experiment}
The goal of this section is to lower bound $\mu'$. We use a branch and bound procedure to lower $\mu'$ and which gives a guarantee for the approximation factor of our algorithm (Theorem \ref{theorem:finalAlg}) . We recall the definition of $\mu'$ 
\begin{align}
    \mu' \triangleq \inf _{\bc \in \C, g(\bc) \geq \delta'} \left\{ \frac{f(\bc)}{g(\bc)}\right\}.
\end{align}

The following claim gives a simple, yet useful, bound on the probability that our rounding algorithm will separate two vertices in the graph.

\begin{claim}
\label{claim:marginalLB}
    Let $u, v \in V$, let $x_i = \|\by_u^i\|^2$ and $w_i = \|\by_v^i\|^2$ for $i = 1, 2, 3$ and $\bc$ be a configuration corresponding to this pair. Then it holds that
    \[ f(\bc) \geq \frac{|x_1 - w_1| + |x_2 - w_2| + |x_3 - w_3|}{2}.\]
\end{claim}

We characterize the configuration space $\C$ which we will consider in the computer assisted proof. 

\begin{lemma}
\label{lem:feasibleConfigurations}
    Let $\bc = (x_1,x_2,x_3,w_1,w_2,w_3,\alpha_1,\alpha_2,\alpha_3, t_1, t_2, t_3) \in \C$. Then $\bc$ satisfies the following:
    \begin{enumerate}
        \item $x_1 + x_2 + x_3 = w_1 + w_2 + w_3 =1$.
        \item $0 \leq \alpha_i \leq \min\{x_i,w_i\}$ for all $i=1,2,3$.
        \item $\max\{0, x_3 -\alpha_3 +\alpha_1 -w_1, w_2 -\alpha_2 +\alpha_1 -x_1\} \leq \min \{w_2 -\alpha_2, x_3 -\alpha_3, x_2 +x_3 -w_1 +\alpha_1 -\alpha_2 -\alpha_3\}.$
        \item $t_i = \frac{\alpha_i - x_iw_i}{\sqrt{(x_i - x_i^2)(w_i - w_i^2)}}$ for $i \in [3]$.
    \end{enumerate}
\end{lemma}

We define the following two polytopes which we will consider when verifying bounding $\mu'$. These polytopes help us to speed up the branch and bound procedure. 
 \begin{enumerate}
     \item $\mathcal{S} \triangleq \{ (x_1, x_2, x_3, w_1, w_2, w_3,\alpha_1, \alpha_2, \alpha_3, t_1, t_2, t_3) |  x_1 \leq \min(x_2, w_1, w_2, w_3, x_3), x_2 \leq x_3 \}$.
     \item  $\mathcal{E} \triangleq \{ \bc = (x_1, x_2, x_3, w_1, w_2, w_3, \alpha_1, \alpha_2, \alpha_3, t_1, t_2, t_3) \}$ such that 
\begin{align*}
    & \sum_{i = 1}^3 \frac{|x_i - w_i|}{2} \leq \rho g(\bc) , g(\bc) \geq \delta'  \}.
\end{align*}
 \end{enumerate}

The following claim shows why we can restrict to configurations in $\mathcal{S}$ and follows from the symmetry of $f, g$. 
\begin{claim} \label{claim:symmetric}
    Let $\bc \in \mathcal{C} - \mathcal{S}$. Then there exists $\bc' \in \mathcal{C} \cap \mathcal{S}$ such that $f(\bc) = f(\bc')$ and $g(\bc) = g(\bc')$. 
\end{claim}
Then Claim \ref{claim:symmetric} and Claim \ref{claim:marginalLB}  imply the following. 
\begin{claim}\label{claim:experiment_goal}
    If  $ \inf _{\bc \in \C \cap \mathcal{S} \cap \mathcal{E} } \frac{f(\bc)}{g(\bc)} \geq \rho$ then $\mu' \geq \rho$.
\end{claim}
\begin{proof}
    By Claim \ref{claim:symmetric} we get the following: $\mu' = \inf_{\bc \in \C, g(\bc) \geq \delta'} \frac{f(\bc)}{g(\bc)} = \inf_{\bc \in \C \cap \mathcal{S}, g(\bc) \geq \delta'} \frac{f(\bc)}{g(\bc)}$. Assume for contradiction that $\mu' = \min_{\bc \in \C \cap \mathcal{S}, g(\bc) \geq \delta'} \frac{f(\bc)}{g(\bc)} < \rho$. Then $\exists \bc' \in \C \cap \mathcal{S}$ with $g(\bc') \geq \delta'$ and $\frac{f(\bc')}{g(\bc')} < \rho$. We have that $\bc' \notin\mathcal{E}$ otherwise $g(\bc') \geq \rho$ by the assumption of the lemma. Then by definition of $\mathcal{E}$,  $\sum_{i = 1}^3 \frac{|x_i - w_i|}{2} > \rho g(\bc')$, but this is a contradiction by Claim \ref{claim:marginalLB} since $f(\bc') > \sum_{i = 1}^3 \frac{|x_i - w_i|}{2} > \rho g(\bc')$.
\end{proof}
Thus our goal for the computed assisted proof is to show $\inf _{\bc \in \C \cap \mathcal{S} \cap \mathcal{E} } \frac{f(\bc)}{g(\bc)} \geq \rho$ for some value of $\rho$.
Given $12$ intervals $(I_1, \ldots, I_{12})$ we define the following polytopes to divide our feasible region into hypercubes. Consider the following polytope,
\[
     \mathcal{P}(I_1, \ldots, I_{12}) = (I_1 \times I_2 \ldots \times I_{12}) \cap \mathcal{C} .
\]
Note that intervals $I_j$ correspond to possible values of $x_j$ for $j \in [3]$, intervals $I_3, I_4, I_5$ correspond to values of $w_1, w_2, w_3$, intervals $I_7, I_8, I_9$ correspond to values of $\alpha_1, \alpha_2, \alpha_3$, and intervals $I_{10}, I_{11}, I_{12}$ correspond to $t_1, t_2, t_3$. Our computer assisted proof enumerates $I_1, \ldots, I_{12}$ so that the union of all  $\mathcal{P}(I_1, \ldots, I_{12}) \cap \mathcal{S} \cap \mathcal{E}$ covers $\C \cap \mathcal{S} \cap \mathcal{E}$. For each $I_1, \ldots, I_{12}$ we show one of the following three:
\begin{enumerate}
    \item $\mathcal{P}(I_1, \ldots, I_{12}) \cap \mathcal{S} \cap \mathcal{E} = \emptyset$.
    \item $\inf_{\bc \in \mathcal{P}(I_1, \ldots, I_{12}) \cap \mathcal{S} \cap \mathcal{E}} \frac{f(\bc)}{g(\bc)} \geq \rho$.
    \item Divide $(I_1, \ldots, I_{12})$ into a collection $\mathcal{U}$ whose union equals  $(I_1, \ldots, I_{12})$ so that each $(I_1', \ldots, I_{12}') \in \mathcal{U}$  satisfies one of the first two items.
\end{enumerate}
which implies the hypothesis of Claim \ref{claim:experiment_goal}. The third item is the branching step and the first two items are how we eliminate branches. For our computer assisted proof we will only consider $x_1, x_2, w_1, w_2, t_1, t_2, t_3$ as independent variables. The remaining variables $w_3, x_3, \alpha_1, \alpha_2, \alpha_3$ will always take values $w_3 = 1 - w_1 - w_2, x_3 = 1 - x_1 - x_2, \alpha_i = x_i w_i + t_i\sqrt{(x_i - x_i^2)(w_i - w_i^2)}$ for $i \in [3]$. Our algorithm runs in stages. In the first stage we use an LP to eliminate hypercubes. The first stage works as follows, 
\begin{enumerate}
    \item Enumerate all $\mathcal{I} =  (I_1,\ldots, I_{12})$ such that $|I_j| = \eta_1$ for $j \in \{1,2, 4, 5\}$ and $|I_j| = \eta_2$ for $j = 10, 11, 12$ that the union of all $(I_1, I_2, I_3, I_4, I_{10}, I_{11}, I_{12})$  covers the region $[0, 1]^4 \times [-1, 1]^3$. Note that intervals $I_5, I_6, I_7$ can be determined by the bounds on the other intervals, $I_j$ such that $ j \notin \{5, 6, 7 \}$. This follows since bounds on $x_i, w_i, t_i$ imply bounds on $\alpha_i$ because $\alpha_i = t_i \sqrt{(x_i - x_i^2)(w_i - w_i^2)} + x_iw_i$. Similarly, $I_3, I_6$ are determined by $I_1, I_2$ and $I_4, I_5$ respectively since $x_3 = 1- x_1 - x_2$ and $w_3 = 1 - w_1 - w_2 $.
    
    \item For each hypercube $\mathcal{I}$ enumerated in the previous step, check that $\mathcal{P}(\mathcal{I}) \cap   \mathcal{E} \cap \mathcal{S}$ contains a feasible point by solving an LP. If yes, save $\mathcal{I}$ for further processing. 
\end{enumerate}


\paragraph*{Partial Derivatives} 
A crucial ingredient of the branch and bound procedure is to obtain a lower bound on the function $f(\bc)$ for any configuration $\bc\in \mathcal{I}$ in any cube. This we do by computing $f(\bc^*)$ for some well chosen $\bc^*\in \mathcal{I}$ and then using bounds on the partial derivatives to infer a bound $f(\bc)$ for all other $\bc\in \mathcal{I}$. A tight bound on partial derivatives ensures a smaller branch and bound tree. We detail the partial derivatives and bounds thus obtained now.
Previously, we defined the notion of configuration and the function $f$ as a function of $12$ variables. Since we only consider $(x_1, x_2, w_1, w_2, t_1, t_2, t_3)$ as variables we have $\frac{\partial f}{\partial x_3} = \frac{\partial f}{\partial w_3} =  \frac{\partial f}{\partial \alpha_i} = 0$ and $f, g$ are functions of $7$ independent variables. We recall the definition of $f$

\begin{align*}
f(x,w,\alpha,t) = & 1 - \frac{1}{6} \sum_{ \pi \in \mathcal{S}_3} \Bigg[ \Gamma_{t_{\pi(1)}}\left( x_{\pi(1)}, w_{\pi(1)}\right)  + \Gamma_{t_{\pi(1)}}\left( 1- x_{ \pi(1)},1 - w_{\pi(1)}\right) \cdot \\
& \cdot \left(  \Gamma_{t_{\pi(2)}}\left(1 - \frac{x_{\pi(2)}}{1 - x_{\pi(1)}}, 1 - \frac{w_{\pi(2)}}{1 - w_{\pi(1)}}\right) + \Gamma_{t_{\pi(2)}}\left(\frac{x_{\pi(2)}}{1 - x_{\pi(1)}}, \frac{w_{\pi(2)}}{1 - w_{\pi(1)}}\right)  \right)              \Bigg].
\end{align*}
Moreover, for our use we can assume that the domain of $f$ is $0 \leq x_1 +x_2 \leq 1$, $0 \leq w_1 +w_2 \leq 1$, $\alpha_1, \alpha_2, \alpha_3 \in [0,1]$ and $t_1, t_2, t_3 \in [-1,1]$. In the following, we are going to prove bounds on the partial derivatives of $f(x,w,\alpha,t)$ with respect to $x_i$, $w_i$ and $t_i$.

\begin{lemma}\label{lem:main_derivative}
For each $(x,w,\alpha,t)$ in the domain of configuration, the following bounds on the partial derivatives of $f$ hold:
\begin{enumerate}
   \item $ \frac{\partial}{\partial {t_i}}f(x,w,\alpha,t) < 0$, for $i=1,2,3$.
    \item $|\frac{\partial}{ \partial x_i} f(x, w, \alpha, t)| \leq \frac{5}{3}-\frac{1}{3}(1-x_{3-i})$, for $i=1,2$.
    \item $|\frac{\partial}{ \partial w_i} f(x, w, \alpha, t)| \leq \frac{5}{3}-\frac{1}{3}(1-w_{3-i})$, for $i=1,2$.
\end{enumerate}
\end{lemma}

The proof of the lemma is technical and appears in appendix \ref{apx:parDerivProofs}. We now have the following claim that gives a lower bound on all configurations in the hypercube as compared to the point $\mathbf{m}$ in it.

\begin{claim} \label{Claim:midpoint}
     Let $\mathcal{I}$ be a hypercube and $\mathcal{Q} \triangleq \mathcal{P}(\mathcal{I}) \cap \mathcal{S} \cap \mathcal{E}$. Let $\mathbf{m} \in \mathbb{R}^{12}$ be a point where the coordinates corresponding to $x_1, x_2, w_1, w_2$ are the midpoints $I_1, I_2, I_4, I_5$ respectively and the last $3$ coordinates corresponding are $\overline{t_1}, \overline{t_2}, \overline{t_3}$ which are the upper bounds of $I_{10}, I_{11}, I_{12}$. Then the following holds, 
     \begin{align*}
     \min_{ \bc = (x_1, \ldots, t_1, t_2, t_3) \in \mathcal{Q} }  f(x_1, x_2, \ldots, \overline{t_1}, \overline{t_2}, \overline{t_3})  \geq f(\mathbf{m}) - \frac{1}{6} \sum_{z \in \{x, w \}}\sum_{i = 1}^2 (5 - (1 - \overline{z_{3 - i}}))(\overline{z_i} - \underline{z_i}).
    \end{align*}
\end{claim}

\begin{lemma} \label{lemma:intElim}
    Let $\mathcal{I}, \mathbf{m}, \mathcal{Q}$ be defined as in Claim \ref{Claim:midpoint}.  Then $\min_{\bc \in \mathcal{Q}} \frac{f(\bc)}{g(\bc)} \geq \rho$ if,  
    \begin{align*} f(\mathbf{m}) - \frac{1}{6} \sum_{z \in \{x, w \}}\sum_{i = 1}^2 (5 - (1 - \overline{z_{3 - i}}))(\overline{z_i} - \underline{z_i}) 
    \geq  \rho \max_{ \bc \in \mathcal{Q}} g(\bc) \end{align*}
    where $\overline{z_i},\underline{z_i}$ for $z \in \{w, x\}$ are the upper and lower bounds given by their corresponding interval in $\mathcal{I}$.
\end{lemma}
\begin{proof}
    We use the fact that $f$ is non-increasing in $t_i$  and Claim \ref{Claim:midpoint} to get,  
    \begin{align*}
        \min_{ \bc \in \mathcal{Q} }  f(\bc)   & \geq   \min_{ \bc= (x_1, x_2, \ldots, t_1, t_2, t_3) \in \mathcal{Q} }  f(x_1, x_2, \ldots, \overline{t_1}, \overline{t_2}, \overline{t_3}) \\ & \geq
        f(\mathbf{m}) - \frac{1}{6} \sum_{z \in \{x, w \}}\sum_{i = 1}^2 (5 - (1 - \overline{z_{3 - i}}))(\overline{z_i} - \underline{z_i})
    \end{align*}   
    The first inequality follows since  $f$ is non-increasing in $t, \alpha$ by Lemma \ref{lem:main_derivative}. The third inequality follows by Claim \ref{Claim:midpoint}. Thus by the inequality in the lemma and the relation above, we have shown:
         $ \min_{ \bc \in \mathcal{Q} }  f(\bc)   \geq \rho \max_{ \bc \in \mathcal{Q}} g(\bc)$. 
\end{proof}

Now we describe the final stage of the experiment which eliminates all remaining cubes from the first stage. 
\begin{enumerate}
    \item For all remaining hyper cubes $\mathcal{I}$ from stage 1 run the following steps until all cubes are eliminated. 
    \item Split the intervals $I_1, I_2, I_4, I_5$ corresponding to $x_1, x_2, w_1, w_2$ into halves to get $16$ smaller sub-hypercubes $(I_1', \ldots , I_9',  I_{10}, I_{11}, I_{12})$. We note that this also tightens the intervals corresponding to $x_3, w_3, \alpha_1, \alpha_2, \alpha_3$. Check if each smaller hypercube has a feasible point in $\mathcal{P}(I_1', \ldots, I_9', I_{10}, I_{11}, I_{12}) \cap \mathcal{S} \cap \mathcal{E}$.  If yes, then the sub-hypercube can be eliminated. 
    \item  Otherwise, verify if the inequality in \ref{lemma:intElim} holds. If yes, the sub-hypercube can be eliminated. 
    \item Otherwise, split the $t_1, t_2, t_3$ intervals into halves to get $8$ sub-hypercubes \\ $\mathcal{I}'= (I_1', \ldots, I_9', I_{10}'', I_{11}'', I_{12}'')$. This also tightens the intervals for $\alpha_1, \alpha_2, \alpha_3$. Check if each smaller hypercube $\mathcal{I}'$ has a feasible point in $\mathcal{P}(\mathcal{I}') \cap \mathcal{S} \cap \mathcal{E}$.  If yes, then the sub-hypercube can be eliminated. 
    \item Otherwise, split the sub-hyper cube into $16$ smaller hyper-cubes using Step 2. Repeat Steps 2-5 until all  sub cubes are eliminated. 
\end{enumerate}
We ran this branch and bound procedure with $\rho = 0.80, \delta' = 0.01$ giving a final approximation of $0.795$.

\section{Max-k-Section}
\label{sec:numerAndMaxkSec}
Our algorithm for Max-3-Section can be generalized for larger number of sections in the following way. For a natural number $k\geq 4$, one can write a similar SDP formulation, where each node $v\in V$ has $k$ vectors $\by_v^1,\ldots,\by_v^k$, with similar constraints and objective for the SDP for Max-3-Section: $\text{maximize } \sum_{\{u,v\}\in E} \cdot (1-\sum_{i=1}^k \by_u^i \cdot \by_v^i)$. The rounding algorithm will work in a similar way. For example, we consider $k=4$. We draw a random permutation $\pi$ in $\mathcal{S}^4$ and three random gaussian vectors  $\bg_1, \bg_2, \bg_3$ with coordinates independently distributed by $N(0,1)$. Then, we define $S_{\pi(1)}$ and $S_{\pi(2)}$ in the same way like in the algorithm for \maxtris. Next, we define
\[S_{\pi(3)}  \triangleq \left\{ u \in V : \bz_u^{\pi(3)} \cdot \bg_3 \geq \Phi^{-1}\left(1- \frac{\|\by_u^{\pi(3)}\|^2}{1-\|\by_u^{\pi(1)}\|^2 - \|\by_u^{\pi(2)}\|^2}    \right) \right\}\]
and the four clusters in the output will be $C_{\pi(1)} = S_{\pi(1)}$, $C_{\pi(2)}= S_{\pi(2)} \setminus S_{\pi(1)}$, $C_{\pi(3)}= S_{\pi(3) }\setminus (S_{\pi(2)}\cup S_{\pi(1)})$ and $C_{\pi(4)} = V\setminus (S_{\pi(1)} \cup S_{\pi(2)} \cup S_{\pi(3)})$.
Then, for any constant $k$, we can claim that with high probability the solution is concentrated and can be re-balanced, similarly to Lemma \ref{lem:concentration} and Lemma \ref{lemma:balancing}.

Our goal now is to bound the approximation factors that these algorithms achieve, for each $k$. As we discussed in the previous sections, computing the worst approximation guarantee, or even bounding it analytically is not an easy task. For $k=3$, we used the branch and bound algorithm and presented a lower bound on the approximation ratio. For larger values of $k$, the computer-assisted proof method becomes computationally harder. However, one possible approach is to try and give a numerical estimation for the approximation factor. We will do that in the following way: for each $k$, one can write an optimization problem over $k^2$ variables that represent the inner products $\by_u^i \cdot \by_v^j$, for each $i,j \in [k]$, or as we denoted before, a feasible configuration. Then, we wish to minimize the ratio of the probability that $u,v$ are separated and the contribution of the edge $\{u,v\}$ to the SDP, which is $1- \sum_{i=1}^k \by_u^i \cdot \by_v^i$. To solve that optimization problem, we use Matlab and the \textit{fmincon} functionality which can find a local minimum for this optimization problem, but only a local minimum. Therefore, we repeat the experiment for numerous random starting points in the feasible region. The results of the numerical estimations of the approximation for $k=3,4,5$ are presented in Conjecture \ref{conj:final2}.

We note that given a configuration of vectors $\by_u^1,\by_u^2,\by_u^3,\by_v^1,\by_v^2,\by_v^3$ that has a ratio of $\rho$ between the separation probability and the contribution of the edge $(u,v)$ to the SDP solution for Max-3-section, one can construct a configuration for max-4-section by adding two zero vectors $\by_u^4, \by_v^4$. That configuration will have the same contribution for the SDP for max-4-section, and the separation probability will also be the same, even though the algorithm admits four sections. Therefore, in that way of analysis, the approximation ratio of our algorithm can only decrease as $k$ increases. However, our numerical estimations show that for $k\leq 5$, the approximation is not worse than $0.8192$. In addition, we note that the simple algorithm that returns a random balanced $k$-partition achieves a $1-\frac{1}{k}$ approximation, hence for $k=3,4,5$ our algorithm surpasses it.


\bibliography{main}
\appendix

\section{Missing Proofs from Section~\ref{sec:sdp}}
\begin{proof}[Proof of Lemma~\ref{lem:SDP_sum_vectors}]
Consider the product $(\by_u^1+\by_u^2+\by_u^3)\cdot(\by_u^1+\by_u^2+\by_u^3)$.
It follows from Constraints \ref{SDP:distribution} and \ref{SDP:orthogonal} that this product equals $1$.
Thus, $\by_u^1+\by_u^2+\by_u^3$ is a unit vector.
Moreover,  Constraints \ref{SDP:marginal} and \ref{SDP:distribution} give that $\by_{\emptyset}\cdot(\by_u^1+\by_u^2+\by_u^3)=1$.
Since $\by_{\emptyset}$ is also a unit vector (Constraint \ref{SDP:unit_vector}), they must be the same vector.
\end{proof}

\section{Missing Proofs from Section~\ref{sec:algorithm}}
\label{apx:proofsSec4}

\begin{proof}[Proof of Claim \ref{claim:bivariate_gaussian_gamma}]
We observe that $\Pr[X \geq \Phi^{-1}(1 - q_1), Y \geq \Phi^{-1}(1 - q_2)] = \Pr[X \leq -\Phi^{-1}(1 - q_1), Y \leq -\Phi^{-1}(1 - q_2)]$ since $(X,Y)$ has the same distribution as $(-X, -Y)$.
The fact that $-\Phi^{-1}(x) = \Phi^{-1}(1-x)$, $ \forall x\in [0,1]$, concludes the proof.
\end{proof}

\begin{proof}[Proof of Lemma \ref{lemma:approx-final}]
    For each edge $e = \{u, v\}$ in the graph let $\bc_e$ denote the configuration corresponding to that edge determined by the SDP solution $Y$. Let $E_s = \{ e \in E | g(\bc_e) \geq \delta' \}$. For a set $S \subseteq E$ we denote $g(S) = \sum_{e \in S}g(\bc_e)$ and $f(S) = \sum_{e \in S}f(\bc_e) $. Then $g(E) = \mathrm{SDP}_{\mathrm{VAL}}$ and $f(E) = \EE[|\delta(C_1, C_2)| + |\delta(C_1, C_3)| + |\delta(C_2, C_3)|] $.

    Now we show that $| \overline{E_s} | \leq \frac{|E|}{3(1 - \delta')}$. This follows since $g(E) \geq \frac{2}{3}|E|$ so we have, 
    \begin{align*}
        & \frac{2}{3}|E|  \leq g(E) =  g(E_s) + g(\overline{E_s}) 
     \leq |E_s| + \delta' | \overline{E_s} | = |E| - (1 - \delta')| \overline{E_s} |.
    \end{align*}
    The second inequality follows since $g(\bc_e) \leq 1$ for all $e \in E$ and $g(\bc_e) < \delta'$ for $e \notin E_s$. 
    Now we conclude the proof using this fact since
    \begin{align*}
        f(E) &\geq f(E_s) \\
        &\geq \mu' g(E_s) = \mu' (g(E) - g(\overline{E_s})) \\
        &\geq \mu' (g(E) - \delta' |\overline{E_s}| ) \geq \mu' \left(g(E) - \delta' \frac{|E|}{3(1 - \delta')} \right)  \\
        &\geq  \mu' \left(g(E) - \delta' \frac{g(E)}{2(1 - \delta')} \right) = \mu' \left(1 - \frac{\delta'}{2(1 - \delta')}\right) \mathrm{SDP}_{\mathrm{VAL}} .
    \end{align*}
    The second inequality follows by the definition of $\mu'$, the fourth inequality follows from the upper bound on the size of $\overline{E_s}$, and the last inequality follows since $\frac{2}{3}|E|  \leq g(E)$.
\end{proof}

\begin{proof}[Proof of Lemma~\ref{lem:concentration}] Fix $i$ and let $R_u^i$ denote the random variable that indicates that vertex $u$ is assigned to part $i$. Let $P^i\triangleq \frac{1}{n} \sum_{u \in V} R^i_u$. We have $|C_i|=n P^i$ where $C_i$ is the set of vertices assigned to part $i$ by Algorithm \ref{Algorithm:Max3SecAlg}.

For every $i=1,2,3$ and $v\in V$, it follows from Lemma \ref{lem:marginals} that $\Pr[v\in C_i]=\Pr[R_v^i=1]=\|\by_v^i\|_2^2$.
Hence,
\begin{align*}
    \mathbb{E}[P^i]= \frac{1}{n}\sum_{v\in V} \mathbb{E}[R^i_v]=\frac{1}{n} \sum_{v \in V} \|\by_v^i\|_2^2 = \frac{1}{3},
\end{align*}
where the last equality follows from Constraint (\ref{SDP:balance}) of $\mathrm{SDP}$. 
Let us now bound the variance of $ P^i$:
\begin{align}
\mathrm{Var}[P^i] &=  \frac{1}{n^2} \left( \sum_{u \in V} \mathrm{Var}[R_i^u] + 2 \sum_{u \neq v} \mathrm{Cov}(R_i^u, R_i^v)        \right) \leq \frac{1}{3n} + \frac{2}{n^2} \sum_{u \neq v} \mathrm{Cov}(R_u^i, R_v^i).\label{VarBound}
\end{align}
The inequality in (\ref{VarBound}) follows since: 
\begin{align*}
   \sum_{u \in V}\mathrm{Var}[R_i^u] &= \sum_{u \in V} \mathbb{E}[R_i^u] - \sum_{u \in V} \mathbb{E}[R_i^u]^2 
    \leq  \sum_{u \in V} \mathbb{E}[R_i^u] = \sum_{u \in V} \| \by_u^i \|^2 = \frac{n}{3},
\end{align*}
where the last equality above follows again from Constraint (\ref{SDP:balance}) of $\mathrm{SDP}$.
Let us now focus on the second term in (\ref{VarBound}): 
\begin{align*}
    \mathrm{Cov}(R^i_v,R^i_u)=\Pr[R^i_v=1,R^i_u=1]-\Pr[R^i_v=1] \cdot \Pr[R^i_u=1].
\end{align*}
We require the following claim which follows from \cite{AustrinBG16}.
\begin{claim} \label{claim:gamma_bound}
Let $ (X,Y)$ be a standard bi-variate gaussian with correlation $\rho$.
Then for every $ q_1,q_2\in [0,1]$:
$$ \Gamma_{\rho}(1-q_1,1-q_2) \leq  (1-q_1)(1-q_2)+ \min\{2|\rho|, q_1(1 - q_2), q_2(1 - q_1) \}.$$
\end{claim}
\begin{proof}
    We have that $\Gamma_{\rho}(1-q_1,1-q_2) \leq (1 - q_1)(1-q_2) + 2|\rho|$ which is shown in Lemma 2.7 of \cite{AustrinBG16}. Next we have $\Gamma_{\rho}(1-q_1,1-q_2) \leq \min\{1 - q_1, 1 - q_2\}$. Then the claim follows from the fact that $1 - q_2 = (1 - q_1)(1 - q_2) + q_1(1 - q_2)$ and $1 - q_1 = (1 - q_1)(1 - q_2) + q_2(1 - q_1)$.  
\end{proof}

We also require the following claim for our proof.
Recall that for every vertex $u\in V$ and $ i=1,2,3$, the vector $\tilde{\bz}_u^i$ is the unscaled component of $\by_u^i$ in the direction orthogonal to  $ \by_{\emptyset}$, {\em i.e.}, $ \tilde{\bz}_u^i=\sqrt{\| \by_u^i\| ^2 - \| \by_u^i\|^4}\bz_u^i$.
\begin{claim}\label{claim:covbound}
For every $u, v \in V$ and any $i \in [3]$, 
$ \Pr[R_u^i = 1, R_v^i = 1] \leq  x_1 w_1  + \frac{7}{3}\sum_{i = j}^3|\tilde{\bz}_u^j \cdot \tilde{\bz}_v^j|^{1/3} $.
\end{claim}
\begin{proof}
    Without loss of generality, we show the inequality for $i = 1$ and by symmetry the proof follows for $i = 2, 3$. First we condition the permutation $\pi$ and look the following three cases: $\pi(1) = 1, \pi(2) = 1, \pi(3) = 1$. For each case the probability contribution is, 
\begin{enumerate}
    \item $\frac{1}{3}\Gamma_{t_1}(x_1, w_1)$
    \item $\frac{1}{6}\Gamma_{t_2}(1 - x_2, 1 - w_2)\Gamma_{t_1}(\frac{x_1}{1-x_2}, \frac{w_1}{1-w_2})  + \frac{1}{6} \Gamma_{t_3}(1 - x_3, 1 - w_3) \Gamma_{t_1}(\frac{x_1}{1-x_3}, \frac{w_1}{1-w_3})   $ 
    \item $\frac{1}{6}\Gamma_{t_2}(1 - x_2, 1 - w_2) \Gamma_{t_3}(1 - \frac{x_3}{1 - x_2}, 1-\frac{w_3}{1 - w_2} ) + \frac{1}{6} \Gamma_{t_3}(1 - x_3, 1 - w_3) \Gamma_{t_2}(1 - \frac{x_2}{1 - x_3}, 1-\frac{w_2}{1 - w_3}) $ \\ $= \frac{1}{6}\Gamma_{t_2}(1 - x_2, 1 - w_2) \Gamma_{t_3}(\frac{x_1}{1 - x_2}, \frac{w_1}{1 - w_2} ) + \frac{1}{6} \Gamma_{t_3}(1 - x_3, 1 - w_3) \Gamma_{t_2}(\frac{x_1}{1 - x_3}, \frac{w_1}{1 - w_3}) $
\end{enumerate}
Let $R_i, Y_i \sim N(0, 1)$ with $Cov(R_i, Y_i) = t_i$ for $i = 1, 2, 3$. The first case follows since $\pi(1) = 1$ with probability $1/3$ and $u, v$ go in $C_1$  if $R_1 \geq \Phi^{-1}(1-x_1), Y_1 \geq \Phi^{-1}(1-w_1)$. For case two, with $1/6$ probability the $\pi(1) = 2, \pi(2) = 1$. Then for both $u, v$ to go in $C_1$ they cannot go to $S_{\pi(1)} = S_2$ and must go to $S_{\pi(2)} = S_1$ (which are independent events). The vertices $u, v$ do not go to $S_{\pi(1)} = S_2$ if $R_2 
\leq \Phi^{-1}(1-x_2), Y_2 \leq \Phi^{-1}(1-w_1)$ which occurs with probability $\Gamma_{t_2}(1 - x_2, 1 - w_2)$. Then, $u, v$ go to $C_1$ if $X_1 \geq 1 - \frac{x_1}{1-x_2}, Y_1 \geq 1 - \frac{w_1}{1 - w_2}$. A similar calculation gives the probability for when $\pi(1) = 3, \pi(2) = 1$. For the third, case, with $1/6$ probability $\pi(1) = 2, \pi(2) = 3$ and $u, v$ go to $C_1$ if they do not go to both $S_{\pi(1)}, S_{\pi(2)}$ which are independent events. The vertices $u, v$ do not go to $S_{\pi(1)} = S_2$ with probability $\Gamma_{t_2}(1 - x_2, 1-w_2)$ and do not go to $S_{\pi(2)} = S_3$ with probability $\Gamma_{t_3}(1 - \frac{x_3}{1 - x_2}, 1 - \frac{w_3}{1-w_2})$. A similar calculation follows for when $\pi(1) = 3, \pi(2) = 2$. Let $m_i = \min\{2|t_i|,x_i(1-w_i),w_i(1 - x_i)\}$ for $i =1, 2, 3$ and $m_4 = \min\{2|t_1|, \frac{x_1}{1-x_2}\frac{w_3}{1-w_2}, \frac{w_1}{1-w_2}\frac{x_3}{1-x_2}\}$, $m_5 = \min\{2|t_3|, \frac{x_1}{1-x_2}\frac{w_3}{1-w_2}, \frac{w_1}{1-w_2}\frac{x_3}{1-x_2} \}$, $m_6 = \min\{2|t_2|, \frac{x_1}{1-x_3}\frac{w_2}{1-w_3}, \frac{w_1}{1-w_3}\frac{x_2}{1-x_3}  \}$, and $m_7 = \min\{2|t_1|, \frac{x_1}{1-x_3}\frac{w_2}{1-w_3}, \frac{w_1}{1-w_3}\frac{x_2}{1-x_3} \}$.
Applying the upper bounds from Claim \ref{claim:gamma_bound} on $\Gamma$ each term is at most
\begin{enumerate}
    \item $\frac{1}{3}( x_1 w_1 + m_1)$
    \item $\frac{1}{6}\left( \left( (1-x_2)(1-w_2) + m_2 \right) \left(\frac{x_1 w_1}{(1-x_2)(1-w_2)} + m_4\right)\right) + \frac{1}{6} \left( \left( (1 - x_3)(1 - w_3)  + m_3 \right) \left( \frac{x_1 w_1}{(1- x_3)(1 - w_3)} + m_7  \right)  \right) $
    
    \item $\frac{1}{6} \left(  \left( (1-x_2)(1-w_2) + m_2  \right)  \left(  \frac{x_1 w_1}{(1-x_2)(1-w_2)}  + m_5 \right)  \right) + \frac{1}{6} \left(  \left( (1-x_3)(1-w_3) + m_3  \right)  \left(   \frac{x_1 w_1}{(1-x_3)(1-w_3)}  + m_6 \right)  \right)$. 
\end{enumerate}
We show the following four bounds which we use to simplify the above expression, 
\begin{enumerate}
    \item $(1-x_2)(1 - w_2)m_4 \leq m_1$
    \item $(1-x_2)(1 - w_2)m_5 \leq m_3$
    \item $(1-x_3)(1 - w_3)m_6 \leq m_2$
    \item $(1-x_3)(1 - w_3)m_7 \leq m_1$.
\end{enumerate}
We observe that $(1-x_2)(1 - w_2)m_4 = \min\{2|t_1|(1-x_2)(1 - w_2), x_1w_3, w_1x_3 \} \leq \min\{2|t_1|, x_1(1 - w_1), w_1(1-x_1)\} = m_1 $ which follows since $(1-x_2)(1 - w_2)\leq 1, w_3 \leq 1 - w_1$, and $x_3 \leq 1- x_1$. A similar calculation can be used to show the remaining three items. Applying these upper bounds and summing the three cases we get
\begin{align*}
    &x_1w_1 + \frac{m_1}{2} + \frac{m_2}{6}\frac{x_1w_1}{(1 - x_2)(1 - w_2)} + \frac{m_2m_4}{6} + \frac{m_1}{6} + \frac{m_3}{6}\frac{x_1w_1}{(1 - x_3)(1 - w_3)} + \\
    &\frac{m_3m_7}{6} + \frac{m_3}{6} + \frac{m_2}{6}\frac{x_1w_1}{(1 - x_2)(1 - w_2)} + \frac{m_2m_5}{6} + \frac{m_2}{6}  + \frac{m_3}{6}\frac{x_1w_1}{(1 - x_3)(1 - w_3)} + \frac{m_3m_6}{6}.
\end{align*}
Then using the fact that $x_i/(1 - x_j) \leq 1$ and $w_i/(1 - w_j) \leq 1$ for $i \neq j$ and simplifying we get that the last expression is at most
\begin{align*}
    x_1w_1 + \frac{2m_1}{3} + \frac{m_2}{2} + \frac{m_3}{2} + \frac{m_2m_4}{6} + \frac{m_3m_7}{6}  + \frac{m_2m_5}{6}   + \frac{m_3m_6}{6} \leq x_1w_1 + \frac{2m_1}{3} + \frac{7}{6}m_2 + \frac{7}{6}m_3.
\end{align*}
The last inequality follows since $m_4 \leq 2|t_1| \leq 2$, $m_7 \leq 2$, $m_5 \leq 2$, and $m_6 \leq 2$. We now show that for $i \in [3]$, $m_i \leq 2|  \tilde{\bz}_u^i \cdot  \tilde{\bz}_v^i  |^{1/3}$ which will finish the proof. We use the fact that $|  \tilde{\bz}_u^i \cdot  \tilde{\bz}_v^i  | = |t_i| \sqrt{x_i(1 - w_i)}\sqrt{w_i(1-x_i)} $ which implies that at least one of the following holds
\begin{enumerate}
    \item $|t_i| \leq |  \tilde{\bz}_u^i \cdot  \tilde{\bz}_v^i  |^{1/3}$
    \item $x_i(1 - w_i) \leq | \tilde{\bz}_u^i \cdot  \tilde{\bz}_v^i  |^{2/3} $
    \item $w_i(1 - x_i ) \leq | \tilde{\bz}_u^i \cdot  \tilde{\bz}_v^i  |^{2/3}. $
\end{enumerate}
Since one of these holds and $m_i = \min\{2|t_i|, x_i(1-w_i), w_i(1-x_i) \}$ we get that 
\[ m_i \leq \max\{2|\tilde{\bz}_u^i \cdot  \tilde{\bz}_v^i  |^{1/3}, | \tilde{\bz}_u^i \cdot  \tilde{\bz}_v^i  |^{2/3}    \} = 2|  \tilde{\bz}_u^i \cdot  \tilde{\bz}_v^i  |^{1/3} \] where the last inequality follows since $0 \leq | \tilde{\bz}_u^i \cdot  \tilde{\bz}_v^i  | \leq 1$. 
\end{proof}

Using the above we prove the following:

Now Claim \ref{claim:covbound} implies that: 
\begin{align*}
    \mathrm{Cov}(R^i_v,R^i_u)&=\mathbb{E}[R^i_v,R^i_u]-\mathbb{E}[R^i_v]\cdot \mathbb{E}[R^i_u]\\
    &\leq \|\by_u^i\|_2^2 \|\by_v^i\|_2^2 +  \frac{7}{3}\sum_{i = 1}^3 |\langle \tilde{\bz}_u^i, \tilde{\bz}_v^i \rangle|^{1/3} - \|\by_u^i\|_2^2 \|\by_v^i\|_2^2 \\
    &=   \frac{7}{3}\sum_{i = 1}^3 |\langle \tilde{\bz}_u^i, \tilde{\bz}_v^i \rangle|^{1/3} .
\end{align*}

Now, Lemma 5.5 \cite{RaghavendraT12} implies the following claim in our notation. 

\begin{claim} \label{cl:itbound}
For all $u \neq v \in V$, we have that $\mathrm{Cov}(R_u^i, R_v^i) \leq 8 I_{\mu_{u,v}}(X_u,X_v)^{1/6}. $
\end{claim}
\begin{proof}
    By Fact 4.3 in \cite{RaghavendraT12} we have that for any $i \in [3]$, 
    \[ I_{\mu_{u,v}}(X_u,X_v) \geq \frac{1}{2} (\Pr[X_u = i, X_v = i] - \Pr[X_u = i]\Pr[X_v = i])^2 = \frac{1}{2}(\by_u^i \cdot \by_v^i  - \|\by_u^i  \| \|\by_v^i  \|^2   )^2 = \frac{1}{2}(\tilde{\bz}_v^i \cdot \tilde{\bz}_u^i )^2 .  \] Then by raising each side to the the power $1/6$ and taking the average over $i = 1, 2, 3$ we get
    \begin{align*}
        I_{\mu_{u,v}}(X_u,X_v)^{1/6} &\geq \frac{1}{3\times 2^{1/6}} \sum_{i = 1}^3 |\langle \tilde{\bz}_u^i, \tilde{\bz}_v^i \rangle|^{1/3} \\
        &\geq \frac{1}{7\times 2^{1/6}}  \mathrm{Cov}(R^i_v,R^i_u).
    \end{align*}
The second inequality follows by Claim \ref{claim:covbound}.
\end{proof}

Now, using the fact that $Y$ is $\frac{1}{t}$-independent we obtain that 
$$\EE_{u,v}[I_{\mu_{u,v}}(X_u,X_v)]\leq \frac{1}{t} ,$$ 
where $\mu_{u,v}$ is the local distribution as given by the SDP solution. 

Now, as in ~\cite{RaghavendraT12}, we obtain that 

\begin{align*}
    \mathrm{Var}[P^i] &\leq \frac{1}{3n} + \frac{2}{n^2} \sum_{u \neq v} \mathrm{Cov}(R_u^i, R_v^i) \\
    &\leq \frac{1}{3n} + \frac{16}{n^2} \sum_{u \neq v}  I_{\mu_{u,v}}(X_u,X_v)^{1/6} \\ 
    &=  \frac{1}{3n} + 16 \EE_{u, v}[ I_{\mu_{u,v}}(X_u,X_v)^{1/6}] \\
    &\leq \frac{1}{3n} + 16  \EE_{u, v}[ I_{\mu_{u,v}}(X_u,X_v)]^{1/6}  \\
    &\leq  \frac{1}{3n}  + 16\frac{1}{t^{1/6}} = O(\varepsilon^{3})
\end{align*}
The third inequality follows since $x^{1/6}$ is concave and the last inequality follows since $t = \Omega(\varepsilon^{-18})$. Thus by applying Chebyshev's Inequality we have that $\Pr[||C_i| - \frac{n}{3}| \geq \frac{\varepsilon n}{3}] = \Pr[|P^i - \frac{1}{3}| \geq \frac{\varepsilon}{3}]  \leq \frac{\mathrm{Var}[P^i]}{\varepsilon^2/9} \leq \varepsilon $.
\end{proof}

\begin{proof}[Proof of Lemma \ref{lemma:balancing}]
Without loss of generality we will assume $|C_1| < |C_2| \leq |C_3|$. We split the analysis into two cases and in the first case both $C_2, C_3$ will have excess vertices. We have, 
\begin{align*}
    |C_1| &= \frac{n}{3} - m_1 & 0 \leq m_1 \leq n\epsilon/3 \\
    |C_2| &= \frac{n}{3} + e_2 & 0 \leq e_2 \leq n \epsilon/3 \\
    |C_3| &= \frac{n}{3} + e_3 & 0 \leq e_2 \leq n \epsilon/3 \\
    m_1 &= e_2 + e_3.
\end{align*}
In this case we will sample a uniformly random set $S_2 \subseteq C_2$ with $|S_2| = e_2$ and a uniformly random set $S_3 \subseteq C_3$ with $|S_3| = e_3$. Out new cut will be $C_1' = C_1 \cup S_2 \cup S_3, C_2' = C_2 - S_2, C_3' = C_3 - S_3$ which clearly satisfies $|C_1'| = |C_2'| = |C_3'| = \frac{n}{3}$. We observe that the amount of edges lost in the objective value is $|\delta(S_2, C_1)|, |\delta(S_3,C_1)|, |\delta(S_2, S_3)|$. Next we have that $\mathbb{E}[|\delta(S_2, C_1)|] = |\delta(C_2, C_1)| \frac{e_2}{|C_2|}$ and $\mathbb{E}[|\delta(S_3, C_1)|] = |\delta(C_3, C_1)| \frac{e_3}{|C_3|}$ which follows since $S_2, S_3$ are uniformly random sets. To bound the third term we use, $|\delta(S_2, S_3)| \leq |\delta(S_2, C_3)|$ and so $\mathbb{E}[|\delta(S_2, S_3)|] \leq \mathbb{E}[ |\delta(S_2, C_3)|] = |\delta(C_2, C_3)|\frac{e_2}{|C_2|}$. Thus the total expected loss is at most
\[  |\delta(C_3, C_1)| \frac{e_3}{|C_3|}    + \frac{e_2}{|C_2|} 
\left(|\delta(C_2, C_3)| +  |\delta(C_2, C_1)| \right) \leq
    \Delta \left( \frac{e_3}{|C_3|} + \frac{e_2}{|C_2|}  \right)     \] which follows since $|\delta(C_3, C_1)| \leq \Delta$ and $|\delta(C_2, C_3)| +  |\delta(C_2, C_1)|  \leq \Delta$. Then we need to maximize, 
\[ \frac{e_3}{|C_3|} + \frac{e_2}{|C_2|} = \frac{|C_3| - n/3}{|C_3|} + \frac{|C_2| - n/3}{|C_2|} = 2 - \frac{n}{3} \left( \frac{1}{|C_1|} + \frac{1}{|C_2|} \right).    \] Using Jensen's inequality and $m_1 \leq n\epsilon/3$  we conclude this case since $\frac{1}{2}\left(\frac{1}{|C_2|} + \frac{1}{|C_3|}\right) \geq \frac{2}{|C_2| + |C_3|} = \frac{2}{\frac{2n}{3} + m_1} \geq \frac{6}{n(2 + \epsilon)}$ implying that the loss is at most $\Delta(2 - \frac{4}{2 + \epsilon} )=  \frac{2\Delta\epsilon}{2 + \epsilon} \leq 2\Delta\epsilon $.

In the second case both $C_1, C_2$ will have missing vertices and $C_3$ will have all the excess. We have, 
\begin{align*}
    |C_1| &= \frac{n}{3} - m_1 & 0 \leq m_1 \leq n\epsilon/3 \\
    |C_2| &= \frac{n}{3} - m_2 & 0 \leq m_2 \leq n \epsilon/3 \\
    |C_3| &= \frac{n}{3} + e_3 & 0 \leq e_3 \leq n \epsilon/3 \\
    e_3 &= m_1 + m_2.
\end{align*}
In this case we first sample a set $S \subseteq C_3$ of size $e_3$ then we sample a uniformly random set $S_2 \subseteq S$ of size $m_2$. We set $C_1' = C_1 - (S - S_2), C_2'= C_2 \cup S_2, C_3' = C_3 -S$. The loss of edges in the solution is given by $\delta(S_2, C_2), \delta(S - S_2, C_1)$. We observe that $|\delta(S_2, C_2)| \leq |\delta(S, C_2)|$ and $|\delta(S - S_2, C_1)| \leq |\delta(S, C_1)|$ so we can bound the expected loss by
\[\EE[|\delta(S_2, C_2)|] + \EE[|\delta(S - S_2, C_1)|] \leq \EE[|\delta(S, C_2)|] + \EE[|\delta(S, C_1)|] 
   = |\delta(C_3, C_2)| \frac{e_3}{|C_3|} + |\delta(C_3, C_1)| \frac{e_3}{|C_3|} \]
and continue with
\[|\delta(C_3, C_2)| \frac{e_3}{|C_3|} + |\delta(C_3, C_1)| \frac{e_3}{|C_3|} 
    \leq \Delta \left(  \frac{e_3}{|C_3|}  \right) = \Delta \left( 1- \frac{n}{3|C_3|} \right) 
    \leq \Delta \left(1 - \frac{1}{1 + \epsilon} \right) = \Delta \frac{\epsilon}{1 + \epsilon} \leq \Delta\epsilon.\]
The first inequality follows since $|\delta(C_3, C_1)| + |\delta(C_3, C_2)| \leq \Delta$, the second inequality follows since $|C_3| \leq \frac{n}{3}(1 + \epsilon)$, and the third inequality follows since $\epsilon \geq 0$.  
\end{proof}


\section{Proofs of Partial Derivatives from Section~\ref{Section:Experiment}}
\label{apx:parDerivProofs}

The following derivative is shown by \cite{DreznerIntegral}.
\begin{lemma}\label{lemma:partial_ti}
For $q_1, q_2 \in [0, 1]$ and $\rho \in [-1, 1]$ we have, 
\[ \frac{\partial}{\partial \rho} \Gamma_{\rho}(q_1, q_2) = \frac{1}{2\pi \sqrt{1 - \rho^2}} e^{-\frac{ \Phi^{-1}(q_2)^2 + \Phi^{-1}(q_1)^2 - 2\Phi^{-1}(q_2) \Phi^{-1}(q_1) \rho  }{2(1-\rho^2)}}. \]
\end{lemma}

\begin{lemma}\label{lem:partial}
For any $(x,w,\alpha,t)$ in the domain of $f$ , we have that
 $$ \frac{\partial}{\partial{t_i}}f(x,w,\alpha,t) < 0$$ for each $i=1,2,3$.
\end{lemma}
\begin{proof}
Lemma \ref{lemma:partial_ti} shows  $\frac{\partial}{\partial t_i}\Gamma_{t_i}(q_1, q_2)$ is non-negative. In addition, we have that $\frac{\partial}{\partial t_j}\Gamma_{t_i}(q_1, q_2) = 0$ for $j \ne i$. 
We notice that $f(x,w,\alpha,t)$ has the following three types of terms: (1) $\Gamma$ terms with a negative sign, (2) products of two $\Gamma$ terms with negative sign and (3) a constant. Since $\Gamma$ terms are positive (they are probabilities) and have non-negative derivatives with respect  to $t_i$, the claim follows. 
\end{proof}

Now we compute the partial derivative of the cut probability with respect to $x_i$ and $w_i$. We need the following lemma from \cite{AustrinBG16}.

\begin{lemma} \label{lemma:gammaDq}
Let $\Gamma_{\rho}(q_1, q_2) = \Pr[X \leq \Phi^{-1}(q_1), Y \leq \Phi^{-1}(q_2)]$ where $Cov(X, Y) = \rho$ and $q_1, q_2 \in [0, 1]$ $\rho \in [-1, 1]$. Then we have, 
\[ \frac{\partial}{\partial q_1} \Gamma_{\rho}(q_1, q_2) = \Phi\left( \frac{\Phi^{-1}(q_2) - \rho \Phi^{-1}(q_1)}{\sqrt{1 - \rho^2}} \right). \]
\end{lemma}

Note that it follows from Lemma \ref{lemma:gammaDq} that $\frac{\partial}{\partial q_1} \Gamma_{\rho}(q_1, q_2) \in [0,1]$.

\begin{lemma} \label{lemma:gradBoun}
    For any $i \in \{1,2\}$, we have the following bounds on the partial derivatives:
    \begin{enumerate}
        \item $|\frac{\partial}{ \partial x_i} f(x, w, \alpha, t)| \leq \frac{5}{3}-\frac{1}{3}(1-x_{3-i}).$
        \item $|\frac{\partial}{ \partial w_i} f(x, w, \alpha, t)| \leq \frac{5}{3}-\frac{1}{3}(1-w_{3-i}).$
    \end{enumerate}
\end{lemma}
\begin{proof}
    Without loss of generality, we will take the partial derivative with respect to $x_1$. We will consider all possible permutations $\pi \in \mathcal{S}^3$ separately. We denote $D_{\rho}(q_1, q_2):= \frac{\partial }{\partial q_1} \Gamma_{\rho}(q_1, q_2)$ whose expression is given in Lemma \ref{lemma:gammaDq} and we will only use the fact that $0\leq D_{\rho}(q_1, q_2) \leq 1$ since it is a probability. 
    First, we assume that $\pi = (1, 2, 3)$. That is, $\pi(1)=1, \pi(2)=2 , \pi(3)=3$. Then the expression in $f(x,w,\alpha,t)$ that corresponds to $\pi$ is
    \[ \Gamma_{t_1}(x_1, w_1) + \Gamma_{t_1}(1 - x_1,1 - w_1)\left( \Gamma_{t_2}(\frac{x_2}{1 - x_1}, \frac{w_2}{1 - w_1}) + \Gamma_{t_2}(1 - \frac{x_2}{1 - x_1}, 1 - \frac{w_2}{1 - w_1}) \right). \] 
    and its derivative is given by, 
    \begin{align}
        & D_{t_1}(x_1, w_1) - D_{t_1}(1 - x_1, 1 - w_1) \left( \Gamma_{t_2}(\frac{x_2}{1 - x_1}, \frac{w_2}{1 - w_1}) + \Gamma_{t_2}(1 - \frac{x_2}{1 - x_1}, 1 - \frac{w_2}{1 - w_1}) \right) + \label{ineq1} \\  
        & \frac{x_2 \Gamma_{t_1}(1 - x_1, 1 - w_1)}{(1 - x_1)^2} \left( D_{t_2}(\frac{x_2}{1 - x_1}, \frac{w_2}{1 - w_1})  - D_{t_2}(1 - \frac{x_2}{1 - x_1}, 1 - \frac{w_2}{1 - w_1}) \right). \nonumber
    \end{align}
    We have that the expression in \ref{ineq1} is in the range $ [-2+x_3 ,2-x_3]$. These bounds follows since 
   \begin{enumerate}
       \item $0 \leq D_{t_1}(x_1, w_1) \leq 1$.
       \item $0 \leq D_{t_1}(1 - x_1, 1 - w_1) \left( \Gamma_{t_2}(\frac{x_2}{1 - x_1}, \frac{w_2}{1 - w_1}) + \Gamma_{t_2}(1 - \frac{x_2}{1 - x_1}, 1 - \frac{w_2}{1 - w_1}) \right) \leq 1$.
       \item $-1+x_3 \leq \frac{x_2 \Gamma_{t_1}(1 - x_1, 1 - w_1)}{(1 - x_1)^2} \left( D_{t_2}(\frac{x_2}{1 - x_1}, \frac{w_2}{1 - w_1})  - D_{t_2}(1 - \frac{x_2}{1 - x_1}, 1 - \frac{w_2}{1 - w_1}) \right) \leq 1-x_3$.
   \end{enumerate}
   The first item follows since $D$ is a probability. The second item follows since $\Gamma_{t_2}(\frac{x_2}{1 - x_1}, \frac{w_2}{1 - w_1}) + \Gamma_{t_2}(1 - \frac{x_2}{1 - x_1}, 1 - \frac{w_2}{1 - w_1}) \in [0, 1]$ since by definition of $\Gamma$ these are the sums of disjoint probability events and therefore are also probabilities. For upper bound of the third item we have $\frac{x_2}{(1 - x_1)^2}\Gamma_{t_1}(1 - x_1, 1 - w_1) \leq \frac{x_2}{1-x_1}\leq 1-x_3$ where the first inequality follows since  $\Gamma_{t_1}(1 - x_1, 1 - w_1) \leq 1  - x_1$  and the second inequality follows since $\frac{x_2}{x_2+x_3}=1-\frac{x_3}{x_2+x_3} \leq 1-x_3$. The lower bound in the third item follows since $ D_{t_2}(\frac{x_2}{1 - x_1}, \frac{w_2}{1 - w_1})  - D_{t_2}(1 - \frac{x_2}{1 - x_1}, 1 - \frac{w_2}{1 - w_1}) \geq -1$ and $0 \leq \frac{x_2 \Gamma_{t_1}(1 - x_1, 1 - w_1)}{(1 - x_1)^2} \leq 1-x_3$.

   Next, we consider the permutation $\pi = (1, 3, 2)$. Since $x_3$ and $ w_3$ are not in the configuration, we use $x_3 = 1 - x_1 - x_2$ and $w_3=1-w_1-w_2$. Hence the expression that corresponds to $\pi$ is 
   \[ \Gamma_{t_1}(x_1, w_1) + \Gamma_{t_1}(1 - x_1,1 - w_1)\left( \Gamma_{t_3}(\frac{1-x_1-x_2}{1 - x_1}, \frac{1-w_1-w_2}{1 - w_1}) + \Gamma_{t_3}(1 - \frac{1-x_1-x_2}{1 - x_1}, 1 - \frac{1-w_1-w_2}{1 - w_1}) \right) \]  which equals to
   \[ \Gamma_{t_1}(x_1, w_1) + \Gamma_{t_1}(1 - x_1,1 - w_1)\left( \Gamma_{t_3}(1-\frac{x_2}{1 - x_1}, 1-\frac{w_2}{1 - w_1}) + \Gamma_{t_3}( \frac{x_2}{1 - x_1}, \frac{w_2}{1 - w_1}) \right).\]
    We observe that this expression is the same as the expression for $\pi = (1, 2, 3)$ except $t_2$ is replaced with $t_3$. Since $t_2$ and $ t_3$ are independent constants in our configuration, the same bounds for expression \ref{ineq1} hold for the derivative in this case.
    
    Next, we consider $\pi = (2, 1, 3)$. The expression in $f$ that corresponds to $\pi$ is
    \[ \Gamma_{t_2}(x_2, w_2) + \Gamma_{t_2}(1 - x_2, 1 - w_2) \left( \Gamma_{t_1}(\frac{x_1}{1 - x_2}, \frac{w_1}{1 - w_2}) + \Gamma_{t_1}(1 - \frac{x_1}{1 - x_2} ,1 - \frac{w_1}{1 - w_2})  \right). \]
    and its partial derivative with respect to $x_1$ is given by 
    \[  \frac{\Gamma_{t_2}(1 - x_2, 1 - w_2)}{ 1 - x_2} \left( D_{t_1}(\frac{x_1}{1 - x_2}, \frac{w_1}{1 - w_2}) - D_{t_1}(1 - \frac{x_1}{1 - x_2}, 1- \frac{w_1}{1 - w_2}) \right). \]
    This derivative is the interval $[-1, 1]$ since $0 \leq \Gamma_{t_2}(1 - x_2, 1 - w_2) \leq 1 - x_2 \leq 1$ and $-1 \leq  D_{t_1}(\frac{x_1}{1 - x_2}, \frac{w_1}{1 - w_2}) - D_{t_1}(1 - \frac{x_1}{1 - x_2}, 1- \frac{w_1}{1 - w_2} \leq 1$ since this is the difference of two probabilities. The same bounds hold for $\pi = (2, 3, 1)$ since the corresponding expression will be
    \[ \Gamma_{t_2}(x_2, w_2) + \Gamma_{t_2}(1 - x_2, 1 - w_2) \left( \Gamma_{t_3}(\frac{x_3}{1 - x_2}, \frac{w_3}{1 - w_2}) + \Gamma_{t_3}(1 - \frac{x_3}{1 - x_2}, 1 - \frac{w_3}{1 - w_2})  \right) \] which equals
    \[ \Gamma_{t_2}(x_2, w_2) + \Gamma_{t_2}(1 - x_2, 1 - w_2) \left( \Gamma_{t_3}(1-\frac{x_1}{1 - x_2}, 1-\frac{w_1}{1 - w_2}) + \Gamma_{t_3}( \frac{x_1}{1 - x_2}, \frac{w_1}{1 - w_2})  \right) \]
    and its derivative is the same as $(2,1,3)$ except $t_1$ is replaced with $t_3$.

    Next we consider the permutation $\pi = (3, 2, 1)$. The expression for $\pi$ is given by
    \[ \Gamma_{t_3}(1 - x_1 - x_2, 1 - w_1 - w_2) + \Gamma_{t_3}(x_1 + x_2, w_1 + w_2) \left( \Gamma_{t_2}(\frac{x_2}{x_1 + x_2}, \frac{w_2}{w_1 + w_2}) + \Gamma_{t_2}(1 - \frac{x_2}{x_1 + x_2}, 1 -\frac{w_2}{w_1 + w_2})  \right) \]
    and the derivative is
    \begin{align*}
        & -D_{t_3}(x_3, w_3) + D_{t_3}(1 - x_3, 1 - w_3)\left( \Gamma_{t_2}(\frac{x_2}{x_1 + x_2}, \frac{w_2}{w_1 + w_2}) + \Gamma_{t_2}(1 - \frac{x_2}{x_1 + x_2}, 1 -\frac{w_2}{w_1 + w_2}) \right) + \\
        & \frac{x_2}{(x_1 + x_2)^2} \Gamma_{t_3}(x_1 + x_2, w_1 + w_2) \left(D_{t_2}(1 - \frac{x_2}{x_1 + x_2}, 1 -\frac{w_2}{w_1 + w_2}) - D_{t_2}( \frac{x_2}{x_1 + x_2}, \frac{w_2}{w_1 + w_2}) \right). 
    \end{align*} 
    This derivative is in the range $[-2+x_1,2-x_1]$. 
    This is true since the following three bounds hold:
    \begin{enumerate}
        \item $0 \leq D_{t_3}(x_3, w_3) \leq 1$.
        \item $0\leq D_{t_3}(1 - x_3, 1 - w_3)\left( \Gamma_{t_1}(\frac{x_2}{x_1 + x_2}, \frac{w_2}{w_1 + w_2}) + \Gamma_{t_1}(1 - \frac{x_2}{x_1 + x_2}, 1 -\frac{w_2}{w_1 + w_2}) \right) \leq 1$.
        \item $-1+x_1 \leq \frac{x_2}{(x_1 + x_2)^2} \Gamma_{t_3}(x_1 + x_2, w_1 + w_2) \left(D_{t_1}(1 - \frac{x_2}{x_1 + x_2}, 1 -\frac{w_2}{w_1 + w_2}) - D_{t_1}( \frac{x_2}{x_1 + x_2}, \frac{w_2}{w_1 + w_2}) \right) \leq 1-x_1$.
    \end{enumerate}
    The proof of these are similar to the three items proven earlier, except $\frac{x_2}{(x_1 + x_2)^2} \Gamma_{t_3}(x_1 + x_2, w_1 + w_2) \leq \frac{x_2}{x_1+x_2} = 1-\frac{x_1}{x_1+x_2}\leq 1-x_1$.
    Finally, we consider the permutation $\pi=(3,1,2)$. Its expression is
    \[ \Gamma_{t_3}(1 - x_1 - x_2, 1 - w_1 - w_2) + \Gamma_{t_3}(x_1 + x_2, w_1 + w_2) \left( \Gamma_{t_1}(\frac{x_1}{x_1 + x_2}, \frac{w_1}{w_1 + w_2}) + \Gamma_{t_1}(1 - \frac{x_1}{x_1 + x_2}, 1 -\frac{w_1}{w_1 + w_2})  \right) \]
    which equals to
    \[ \Gamma_{t_3}(1 - x_1 - x_2, 1 - w_1 - w_2) + \Gamma_{t_3}(x_1 + x_2, w_1 + w_2) \left( \Gamma_{t_1}(1-\frac{x_2}{x_1 + x_2}, 1-\frac{w_2}{w_1 + w_2}) + \Gamma_{t_1}(\frac{x_2}{x_1 + x_2}, \frac{w_2}{w_1 + w_2})  \right) \]
    Hence the derivative will be the same as $(3,2,1)$ except $t_2$ is replaced with $t_1$.
    Let $D(\pi)$ be the derivative of the expression that corresponds to the permutation $\pi = (i, j, k) \in \mathcal{S}^3$. 
    We sum all bounds and get $\sum_{\pi \in \mathcal{S}^3}D(\pi) \leq 10-2x_1-2x_3$ and $\sum_{\pi \in \mathcal{S}^3}D(\pi) \geq -10+2x_1+2x_3$.
    Finally, since
    \[ \frac{\partial}{ \partial x_1} f(x, w, \alpha, t) = -\frac{1}{6} \sum_{\pi \in \mathcal{S}^3} D(\pi),\] it holds that 
    \[|\frac{\partial}{ \partial x_1} f(x, w, \alpha, t)| \leq \frac{5}{3}-\frac{1}{3}(1-x_2).\]
    The rest of the partial derivatives follow from symmetry between $x_1$ and $x_2$ and symmetry between $x_1,x_2$ and $w_1,w_2$.
\end{proof}

\section{Missing Proofs from Section~\ref{Section:Experiment}}\label{app:Experiment}
\begin{proof}[Proof of Claim \ref{claim:marginalLB}]
It follows from Lemma \ref{lem:marginals} that our algorithm preserves marginals. That is, the probability that $u$ goes to cluster $C_i$ is $x_i$. Hence the probability that both $u$ and $v$ go to cluster $C_i$ is at most $\min\{x_i,w_i\}$. Therefore,
\[\Pr \left[ \text{Algorithm \ref{Algorithm:Max3SecAlg} separates $u$ and $v$}\right] \geq 1 - \sum_{i=1}^3 \min\{x_i,w_i\}.\]
Since $\sum_i x_i = \sum_i w_i =1$, it holds that  $1-\sum_i \min\{x_i,w_i\} = \sum_i \max\{x_i,w_i\} - 1$.  Note that $|x_i-w_i|=\max\{x_i,w_i\}-\min\{x_i,w_i\}$, hence $\frac{1}{2}\sum_i |x_i-w_i|$ is the average between $1-\sum_i \min\{x_i,w_i\}$ and $\sum_i \max\{x_i,w_i\}-1$, and that concludes the proof.
\end{proof}

\begin{proof}[Proof of Claim \ref{Claim:midpoint}]
By the Mean Value Theorem, we have for any differentiable function $g: S \to \mathbb{R}$ and points $a, b \in S$ there exits $\lambda \in [0, 1]$ such that  \[ g(b) \geq g(a) - |\nabla g(\lambda a + (1 - \lambda)b) \cdot (b - a) |.  \] We use this fact and set $g$ to $f$,  $b$ to the optimizer of the LHS in the lemma, $S$ to  $\mathcal{I}$, and $a =\mathbf{m}$. Let $\lambda a + (1 - \lambda)b = (x^{\lambda}_1, x^{\lambda}_2,  x^{\lambda}_3, w^{\lambda}_1, w^{\lambda}_2,  w^{\lambda}_3,  \alpha_1^{\lambda}, \alpha_2^{\lambda}, \alpha_3^{\lambda}, \overline{t_1}, \overline{t_2}, \overline{t_3})$.  Then we have
\begin{align*}
    \min_{  \bc = (x_1, \ldots, t_1, t_2, t_3) \in \mathcal{Q} }  f(x_1, \ldots, \overline{t_1}, \overline{t_2}, \overline{t_3}) \geq  f(\mathbf{m}) - \sum_{z \in \{x, w\}} \sum_{i = 1}^2 \left\lvert \frac{\partial f}{\partial z_i} (\lambda a + (1 - \lambda)b) \right\rvert \frac{|\overline{z_i} - \underline{z_i}|}{2}
\end{align*}

which follows $a$ is the midpoint of variables $x_1, x_2, w_1, w_2$ in $\mathcal{I}$ so the distance between $a, b$ in the marginal variables is at most half the length of their respective intervals in $\mathcal{I}$. We note that the partial derivatives with respect to $x_3, w_3$ are $0$ since their values are determined by $x_1, x_2, w_1, w_2$.  By Lemma  \ref{lemma:gradBoun} we have that, 
\[ | \frac{\partial f(x^{\lambda}_1, x^{\lambda}_2, x^{\lambda}_3, w^{\lambda}_1, w^{\lambda}_2, w^{^{\lambda}}_3,  \alpha_1^{\lambda}, \alpha_2^{\lambda}, \alpha_3^{\lambda}, \overline{t_1}, \overline{t_2}, \overline{t_3}) }{\partial z_i} | \leq \frac{5}{3} - \frac{(1 - z^{\lambda}_{3 - i})}{3} \leq  \frac{5}{3} - \frac{(1 - \overline{z_{3 - i}})}{3}.  \]
The last inequality follows since $\mathcal{I}$ is convex so $a + (1 - \lambda)b \in \mathcal{I}$ and the marginal variables lie in their corresponding intervals of $\mathcal{I}$. 
\end{proof}

\begin{proof}[Proof of Lemma \ref{lem:feasibleConfigurations}]
The last item of the Lemma follows immediately from the last item of Definition \ref{definition:feasible_config}. Given $\bc$ our goal is to see if there exist vectors $\by_u^1$, $\by_u^2$, $\by_u^3$, $\by_v^1$, $\by_v^2$, $\by_v^3$ satisfying the first two conditions of Definition \ref{definition:feasible_config}. We define $\beta_1 \ldots \beta_6$ which will represent the inner products of $\by_u^i \cdot \by_v^j$ for $i \neq j$. We note that these variables do not appear in $\bc$ since we will project these variables out (in this Lemma)  since they do not contribute $f$ or $g$ and only affect feasibility of $\bc$. Let $\beta_1 \triangleq  \by_u^1 \cdot \by_v^2$, $\beta_2 \triangleq \by_u^1 \cdot \by_v^3, \beta_3 \triangleq \by_u^2 \cdot \by_v^1, \beta_4 = \by_u^2 \cdot \by_v^3, \beta_5 \triangleq \by_u^3 \cdot \by_v^2, \beta_6 \triangleq \by_u^3 \cdot \by_v^2$. Constraint (\ref{SDP:distribution}) of SDP \ref{eq:SDP} and the corollary of Lemma \ref{lem:SDP_sum_vectors} give us constraints on the $x, w, \alpha, \beta$ variables. 
Consider 
$Q=\{ ( x_1,x_2,x_3,w_1,w_2,w_3,\alpha_1,\alpha_2,\alpha_3,\beta_1,\beta_2,\beta_3,\beta_4,\beta_5,\beta_6)
\}$ the set of vectors in $\RR^{15}$ that satisfy the following constraints:

\begin{align*}
    &w_1 = \alpha_1 + \beta_3 + \beta_5 &x_1 = \alpha_1 + \beta_1 + \beta_2 \\
    &w_2 = \beta_1 + \alpha_2 + \beta_6 &x_2 = \beta_3 + \alpha_2 + \beta_4 \\
    &w_3 = \beta_2 + \beta_4 + \alpha_3 &x_3 = \beta_5 +\beta_6 +\alpha_3\\
   &\sum_{i=1}^3 x_i = \sum_{i=1}^3 w_i =1     &\alpha_1, \alpha_2, \alpha_3, \beta_1, \beta_2,\beta_3,\beta_4,\beta_5, \beta_6 \geq 0.
\end{align*}

We define $P$ to be the projection of $Q$ on the first $9$ coordinates. That is, a vector \\ $(x_1,x_2,x_3,w_1,w_2,w_3,\alpha_1,\alpha_2,\alpha_3) \in P$ if and only if there exists $\beta_1,\dots,\beta_6 \in \RR$ such that \\ $(x_1,x_2,x_3,w_1,w_2,w_3,\alpha_1,\alpha_2,\alpha_3,\beta_1,\beta_2,\beta_3,\beta_4,\beta_5,\beta_6)\in Q$. Let $\C[9]$ be the projection of the first $9$ coordinates of $\C$. Note that $\C[9] \subseteq P$, since $Q$ corresponds to the constraints in the SDP relaxation. We present the necessary constraints of $P$ by  eliminating $\beta_1,\dots,\beta_6$.
We start by eliminating $\beta_1$ using $\beta_1 = w_2 - \alpha_2 - \beta_6$ and get
\begin{align*}
    &w_1 = \alpha_1 + \beta_3 + \beta_5 &x_1 = \alpha_1 + (w_2 - \alpha_2 - \beta_6) + \beta_2 \\
    &w_2 - \alpha_2 - \beta_6 \geq 0 &x_2 = \beta_3 + \alpha_2 + \beta_4 \\
    &w_3 = \beta_2 + \beta_4 + \alpha_3 &x_3 = \beta_5 +\beta_6 +\alpha_3\\
   &x_1 +x_2 +x_3 = w_1+w_2+w_3 =1     &\alpha_1, \alpha_2, \alpha_3, \beta_2,\beta_3,\beta_4,\beta_5, \beta_6 \geq 0.
\end{align*}
Then, we eliminate $\beta_3$ using $\beta_3 = w_1 - \alpha_1 - \beta_5$ and get
\begin{align*}
    &w_1 - \alpha_1 - \beta_5 \geq 0 &x_1 = \alpha_1 + (w_2 - \alpha_2 - \beta_6) + \beta_2 \\
    &w_2 - \alpha_2 - \beta_6 \geq 0 &x_2 = (w_1 - \alpha_1 - \beta_5) + \alpha_2 + \beta_4 \\
    &w_3 = \beta_2 + \beta_4 + \alpha_3 &x_3 = \beta_5 +\beta_6 +\alpha_3\\
   &x_1 +x_2 +x_3 = w_1+w_2+w_3 =1     &\alpha_1, \alpha_2, \alpha_3, \beta_2,\beta_4,\beta_5, \beta_6 \geq 0.
\end{align*}
Next, we eliminate $\beta_2$ using $\beta_2=x_1-\alpha_1-(w_2-\alpha_2-\beta_6)$
\begin{align*}
    &w_1 - \alpha_1 - \beta_5 \geq 0 &x_1-\alpha_1-(w_2-\alpha_2-\beta_6) \geq 0 \\
    &w_2 - \alpha_2 - \beta_6 \geq 0 &x_2 = (w_1 - \alpha_1 - \beta_5) + \alpha_2 + \beta_4 \\
    &w_3 = x_1-\alpha_1-(w_2-\alpha_2-\beta_6) + \beta_4 + \alpha_3 &x_3 = \beta_5 +\beta_6 +\alpha_3\\
   &x_1 +x_2 +x_3 = w_1+w_2+w_3 =1     &\alpha_1, \alpha_2, \alpha_3, \beta_4,\beta_5, \beta_6 \geq 0.
\end{align*}
We eliminate $\beta_4$ using $\beta_4=x_2 -\alpha_2 - (w_1 - \alpha_1 - \beta_5)$
\begin{align*}
    &w_1 - \alpha_1 - \beta_5 \geq 0 &x_1-\alpha_1-(w_2-\alpha_2-\beta_6) \geq 0 \\
    &w_2 - \alpha_2 - \beta_6 \geq 0 &x_2 -\alpha_2 - (w_1 - \alpha_1 - \beta_5) \geq 0 \\
    &w_3 = x_1-w_2+\beta_6 + x_2 - w_1 + \beta_5 + \alpha_3 &x_3 = \beta_5 +\beta_6 +\alpha_3\\
   &x_1 +x_2 +x_3 = w_1+w_2+w_3 =1     &\alpha_1, \alpha_2, \alpha_3, \beta_5, \beta_6 \geq 0.
\end{align*}
Finally, we eliminate $\beta_5$ using $\beta_5 = x_3- \beta_6 -\alpha_3$ and get
\begin{align*}
    &w_1 - \alpha_1 - (x_3- \beta_6 -\alpha_3) \geq 0 &x_1-\alpha_1-(w_2-\alpha_2-\beta_6) \geq 0 \\
    &w_2 - \alpha_2 - \beta_6 \geq 0 &x_2 -\alpha_2 - (w_1 - \alpha_1 - (x_3- \beta_6 -\alpha_3)) \geq 0 \\
    &w_3 = x_1-w_2 + x_2 - w_1 + x_3 &x_3- \beta_6 -\alpha_3 \geq 0\\
   &x_1 +x_2 +x_3 = w_1+w_2+w_3 =1     &\alpha_1, \alpha_2, \alpha_3, \beta_6 \geq 0.
\end{align*}
To sum up, the conditions that we get are:
\begin{enumerate}
    \item $x_1+x_2+x_3=w_1+w_2+w_3=1$.
    \item $0 \leq a_i \leq \min\{x_i,w_i\}$ for all $i=1,2,3$.
    \item $\max\{0, x_3 -\alpha_3 +\alpha_1 -w_1, w_2 -\alpha_2 +\alpha_1 -x_1\} \leq \min \{w_2-\alpha_2, x_3-\alpha_3, x_2 +x_3 -w_1 +\alpha_1 -\alpha_2 -\alpha_3\}$.
\end{enumerate}
\end{proof}

\begin{observation}
\label{obs:worstConf}
The configuration $\bc=(x_1, x_2, x_3, w_1, w_2, w_3, \alpha_1, \alpha_2, \alpha_3)$ where
\begin{align*}
    &x_1=0.4146,x_2=0.0657,x_3=0.5197,\\
    &w_1=0.0657,w_2=0.4146,w_3=0.5197,\\
    &\alpha_1=0,\alpha_2=0,\alpha_3=0.0393
\end{align*}
admits a ratio of $\frac{f(\bc)}{g(\bc)} \approx 0.8192$.
\end{observation}

\begin{observation}
\label{obs:worstConfNoPerm}
The following configuration admits a ratio of $\approx 0.7192$ between the probability of separation without taking a random permutation on the clusters, and the contribution to the SDP.
\begin{align*}
    &x_1=0.25,x_2=0.25,x_3=0.5,\\
    &w_1=0.25,w_2=0.25,w_3=0..5,\\
    &\alpha_1=0,\alpha_2=0,\alpha_3=0.
\end{align*}
\end{observation}

\end{document}